\documentclass[12pt,notitlepage,a4paper]{article}
\pdfoutput=1

\usepackage{color,graphicx}
\usepackage{amssymb}
\usepackage{delarray,amsmath,bbm}
\usepackage[latin1]{inputenc}
\usepackage[american]{babel}
\usepackage{cite}

\newcommand{\be}{\begin{equation}}
\newcommand{\ee}{\end{equation}}
\newcommand{\beq}{\begin{equation}}
\newcommand{\eeq}{\end{equation}}
\newcommand{\bea}{\begin{eqnarray}}
\newcommand{\eea}{\end{eqnarray}}

\newcommand{\ba}{\begin{eqnarray}}
\newcommand{\ea}{\end{eqnarray}}
\newcommand{\rc}{\nonumber\\}
\newcommand{\bear}{\begin{eqnarray}}
\newcommand{\eear}{\end{eqnarray}}
\def\ie{{\emph{i.e.}}}

\begin{document}
\baselineskip=15.5pt
\pagestyle{plain}
\setcounter{page}{1}

\begin{center}
\vspace{0.1in}

\renewcommand{\thefootnote}{\fnsymbol{footnote}}

\begin{center}
\Large \bf Introduction to the AdS/CFT correspondence
\end{center}
\vskip 0.1truein
\begin{center}
Alfonso V. Ramallo \footnote{alfonso@fpaxp1.usc.es}\\
\end{center}
\vspace{0.5mm}

\begin{center}\it{
Departamento de  F\'\i sica de Part\'\i  culas \\
Universidade de Santiago de Compostela \\
and \\
Instituto Galego de F\'\i sica de Altas Enerx\'\i as (IGFAE)\\
E-15782 Santiago de Compostela, Spain}
\end{center}

\setcounter{footnote}{0}
\renewcommand{\thefootnote}{\arabic{footnote}}

\vspace{0.4in}

\begin{abstract}
\noindent 
This is a pedagogical introduction to the AdS/CFT correspondence, based on lectures delivered by the author at the third IDPASC school. Starting with the conceptual basis of the holographic dualities, the subject is developed  emphasizing some concrete topics, which are discussed in detail. A very brief introduction to string theory is provided, containing  the minimal ingredients to understand the origin of the AdS/CFT duality. Other topics covered are the holographic calculation of correlation functions, quark-antiquark potentials and transport coefficients. 

\end{abstract}

\smallskip
\end{center}

\newpage

\tableofcontents

\section{Introduction and motivation}
\label{sec:1}
The AdS/CFT correspondence is a duality relating quantum field theory (QFT) and gravity. More precisely, the correspondence  relates the quantum physics of strongly correlated many-body systems to the classical dynamics of gravity in one higher dimension. This duality is also referred to as the holographic duality or the gauge/gravity correspondence. In its original formulation \cite{Maldacena:1997re,Gubser:1998bc,Witten:1998qj}, the correspondence related a four-dimensional Conformal Field Theory (CFT) to the geometry of an anti-de Sitter (AdS) space in five dimensions. 

In the study of collective phenomena in condensed matter physics it is quite common that when a system is strongly coupled it reorganizes itself in such a way that new weakly coupled degrees of freedom emerge dynamically and the system can be better described in terms of  fields representing the emergent excitations. The holographic duality is a new example of this paradigm.  The new (and surprising!) feature is that the emergent fields live in a space with one extra dimension and that the dual theory is a gravity theory. As we will argue below, the extra dimension is related to the energy scale of the QFT. The holographic description is a geometrization of the quantum dynamics of the systems with a large number of degrees of freedom, which makes manifest that there are deep connections between quantum mechanics and gravity.

The gauge/gravity duality was discovered in the context of string theory, where it is quite natural to realize (gauge) field theories on hypersurfaces embedded in a higher dimensional space, in a theory containing gravity.  However, the study of the correspondence  has been extended to include very different domains, such as the analysis of the strong coupling dynamics  of QCD and the electroweak theories, the physics of black holes and quantum gravity, relativistic hydrodynamics or different applications in condensed matter physics (holographic superconductors, quantum phase transitions, cold atoms, ...).  In these lectures we will concentrate on some particular topics, trying to present in clear terms the basic conceptual ideas, as well as the more practical calculational aspects of the subject.  For reviews on the different aspects of the duality see \cite{Aharony:1999ti,D'Hoker:2002aw,Hartnoll:2009sz,McGreevy:2009xe,CasalderreySolana:2011us,Kim:2012ey,Adams:2012th}.

\begin{figure}[ht]
\center
\includegraphics[width=0.5\textwidth]{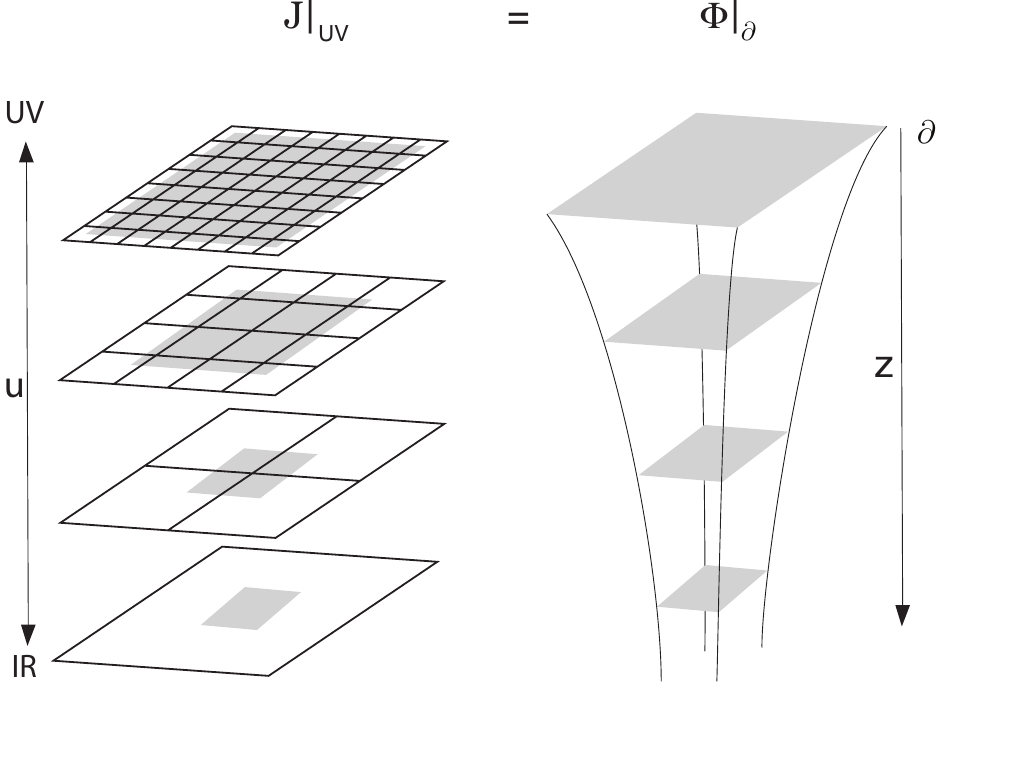}
\caption{On the left we illustrate the Kadanoff-Wilson renormalization of a lattice system. In the AdS/CFT correspondence the lattices at different scale are considered as the  layers of the higher dimensional space represented on the right of the figure.} 
\label{Kadanoff}
\end{figure}

We will start by motivating the duality from the Kadanoff-Wilson renormalization group approach to the analysis of lattice systems. Let us consider a non-gravitational system in a lattice with lattice spacing $a$ and hamiltonian given by:
\beq
H\,=\,\sum_{x,i}\,J_i(x,a)\,\,{\cal O}^i(x)\,\,,
\label{lattice_hamiltonian}
\eeq
where $x$ denotes the different lattice sites and $i$ labels the different operators ${\cal O}^i$. The $J_i(x,a)$ are the coupling constants (or sources) of the operators at the point $x$ of the lattice. Notice that we have included a second argument in $J^i$, to make clear  they correspond to a lattice spacing $a$. In the renormalization group approach we coarse grain the lattice by increasing the lattice spacing and by replacing multiple sites by a single site with the average value of the lattice variables. In this process the hamiltonian retains its form (\ref{lattice_hamiltonian}) but different operators are weighed differently. Accordingly, the couplings $J_i(x,a)$ change in each step. Suppose that we double the lattice spacing in each step. Then, we would have a succession of couplings of the type:
\beq
J_i(x,a)\to J_i(x,2a)\to J_i(x,4a)\to\cdots\,\,.
\eeq
Therefore, the couplings acquire in this process a dependence on the scale (the lattice spacing) and we can write them as $J_i(x,u)$, where $u=(a,2a,4a,\cdots)$ is the length scale at which we probe the system.  The evolution of the couplings with the scale is determined by flow equations of the form:
\beq
u{\partial \over \partial u}\,\,J_i(x,u)\,=\,
\beta_i\Big(J_j(x,u),u\Big)\,\,,
\eeq
where $\beta_i$ is the so-called $\beta$-function of the $i^{\,th}$ coupling constant. At weak coupling the  $\beta_i$'s  can be determined in perturbation theory. At strong coupling the AdS/CFT proposal is to consider $u$ as an extra dimension. In this picture the succession of lattices at different values of $u$ are considered as layers of a new higher-dimensional space. Moreover, the sources $J_i(x,u)$ are regarded as fields in a space with one extra dimension and, accordingly we will simply write:
\beq
J_i(x,u)=\phi_i(x,u)\,\,.
\eeq
The dynamics of the  sources $\phi_i$'s will be governed by some action. Actually, in the AdS/CFT duality the dynamics of the  $\phi_i$'s is determined by some gravity theory ( i. e. by some metric). Therefore, one can consider the holographic duality as a geometrization of the quantum dynamics encoded by the renormalization group. The microscopic couplings of the field theory in the UV can be identified with the values of the bulk fields at the boundary of the extra-dimensional space. Thus, one can say that the field theory lives on the boundary of the higher-dimensional space (see figure \ref{Kadanoff}). 

The sources $\phi_i$ of the dual gravity theory must have the same tensor structure of the corresponding dual operator ${\cal O}^i$ of field theory, in such a way that the product $\phi_i\,{\cal O}^i$ is a scalar. Therefore, a scalar field will be dual to a scalar operator, a vector field $A_{\mu}$ will be dual to a current $J^{\mu}$, whereas a spin-two field $g_{\mu\nu}$ will be dual to a symmetric second-order tensor $T_{\mu\nu}$ which can be naturally identified with the energy-momentum tensor $T_{\mu\nu}$ of the field theory. 

The holographic duality raises several conceptual issues which should be addressed in order to fully understand it. The simplest of these issues is the 
matching of the degrees of freedom on both sides of the correspondence.  Let us consider a QFT in a $d$-dimensional spacetime (\ie\ in $d-1$ spatial dimensions plus time). The number of degrees of freedom of a system is measured by the entropy. On the QFT side the entropy is an extensive quantity. Therefore, if  $R_{d-1}$ is  ($d-1$)-dimensional spatial region, at constant time,  its entropy should be proportional to its volume in $d-1$ dimensions:
\beq
S_{QFT}\propto {\rm Vol}(R_{d-1})\,\,.
\label{QFT_entropy}
\eeq
On the gravity side the theory lives in a $(d+1)$-dimensional spacetime. How such a higher dimensional theory can contain the same information as its lower dimensional dual?. The crucial point to answer this question is the fact  that the entropy in quantum gravity is subextensive. Indeed, in a gravitational theory the entropy in a volume is bounded by the entropy of a black hole that fits inside the volume and, according to the so-called holographic principle, the entropy is proportional to the surface of the black hole horizon (and not to  the volume enclosed by the horizon). More concretely, the black hole entropy is given by the 
Bekenstein-Hawking formula:
\beq
S_{BH}\,=\,{1\over 4 G_N}\,A_{H}\,\,,
\label{BH-formula}
\eeq
where $A_H$ is the area of the event horizon and $G_N$ is the Newton constant. In order to apply (\ref{BH-formula}) for our purposes, let $R_{d}$ be a spatial region in  the $(d+1)$-dimension spacetime where the gravity theory lives and let us assume that  $R_{d}$ is bounded by a $(d-1)$-dimensional manifold $R_{d-1}$ ( $R_{d-1}=\partial R_{d}$). Then, according to (\ref{BH-formula}), the gravitational entropy associated to $R_{d}$ scales as:
\beq
S_{GR} (R_{d})\propto {\rm Area}(\partial R_{d})\propto {\rm Vol} (R_{d-1})\,\,,
\label{GR_entropy}
\eeq
which agrees with the QFT behavior (\ref{QFT_entropy}). In this lectures we will present several refinements of this argument and we will establish a more precise matching of the degrees of freedom of the two dual theories.

\section{The anti-de Sitter space}
\label{sec:2}
In general, finding the geometry associated to a given QFT is a very difficult problem. However, if the theory is at a fixed point of the renormalization group flow (with a vanishing $\beta$-function) it has conformal invariance (is a CFT) and one can easily find such a metric. Indeed,  let us consider a QFT in $d$ spacetime dimensions. 
The most general metric in $(d+1)$-dimensions with Poincar\'e invariance in $d$-dimensions is:
\beq
ds^2\,=\,\Omega^2(z)\,(-dt^2+d\vec x^2+dz^2)\,\,,
\label{CFT-metric}
\eeq
where  $z$  is the coordinate of the  extra dimension, $\vec x=(x^1,\cdots, x^{d-1})$ and 
$\Omega(z)$ is a function to be determined.
If $z$ represents a length scale and the theory is conformal invariant, then $ds^2$ must be invariant under the transformation:
\beq
(t,\vec x)\to \lambda (t,\vec x)\,\,,
\qquad\qquad
z\to \lambda z\,\,.
\label{scale-trans}
\eeq
Then, by imposing  the invariance of  the metric (\ref{CFT-metric}) under the transformation (\ref{scale-trans}), we obtain that the function  $\Omega(z)$ must transform as:
\beq
\Omega(z)\to \lambda^{-1}\,\Omega(z)\,\,,
\eeq
which fixes $\Omega(z)$ to be:
\beq
\Omega(z)={L\over z}\,\,,
\label{Omega}
\eeq
where $L$ is a constant. Thus, by plugging the function (\ref{Omega}) into  the metric (\ref{CFT-metric}) we arrive at the following form of $ds^2$: 
\beq
ds^2\,=\,{L^2\over z^2}\,(-dt^2+d\vec x^2+dz^2)\,\,,
\label{AdS_metric}
\eeq
which is the line element of the $AdS$ space  in $(d+1)$-dimensions, that we will denote by $AdS_{d+1}$.  The constant $L$ is a global factor in (\ref{AdS_metric}), which we will refer to as the anti-de Sitter radius. 
The (conformal) boundary of the $AdS$ space is located at $z=0$. Notice that the metric (\ref{AdS_metric}) is singular at $z=0$. This means that  we will have to introduce a regularization procedure in order to define quantities in the $AdS$ boundary.

The $AdS$ metric (\ref{AdS_metric})  is a solution of the equation of motion of a gravity action of the type:
\beq
I\,=\,{1\over 16\pi G_N}\,
\int d^{d+1}x\,\,\sqrt{-g}\,\Big[\,-2\Lambda\,+\,R\,+c_2\, R^2\,+\,c_3\, R^3\,+\,\cdots]\,\,,
\label{general_gravity_action}
\eeq
with $G_N$ is the Newton constant, the $c_i$ are constants, $g=\det (g_{\mu\nu})$, $R=g^{\mu\nu}R_{\mu\nu}$  is the scalar curvature and $\Lambda$ is a  cosmological constant. In particular, if $c_2=c_3=\cdots = 0$  the action (\ref{general_gravity_action}) becomes the Einstein-Hilbert (EH) action of general relativity with a cosmological constant. In this case the equations of motion  are just  the Einstein equations:
\beq
R_{\mu\nu}\,-\,{1\over 2}\,g_{\mu\nu}\,R\,=\,-\Lambda g_{\mu\nu}\,\,.
\label{Einstein_eq}
\eeq
Taking the trace on both sides of (\ref{Einstein_eq}), we get that the scalar curvature is given by:
\beq
R=g^{\mu\nu}R_{\mu\nu}\,=\,2\,{d+1\over d-1}\,\Lambda\,\,.
\label{Ricci_Cosm_Constant}
\eeq
Inserting back this result in the Einstein equation (\ref{Einstein_eq}) we get that  the Ricci tensor and the metric are proportional:
\beq
R_{\mu\nu}\,=\,{2\over d-1}\,\,\Lambda\,\,g_{\mu\nu}\,\,.
\label{Ricci_eom}
\eeq	
Therefore the  solution of (\ref{Einstein_eq}) defines an Einstein space. Moreover, the Ricci tensor for the metric (\ref{AdS_metric}) can be computed directly from  its definition. We get:
\beq
R_{\mu\nu}\,=\,-{d\over L^2}\,g_{\mu\nu}\,\,.
\label{Ricci_AdS_direct}
\eeq
By comparing  (\ref{Ricci_eom}) and (\ref{Ricci_AdS_direct})  we get that the $AdS_{d+1}$  space solves the equations of motion (\ref{Einstein_eq}) of EH gravity with  a cosmological constant equal to:
\beq
\Lambda\,=\,-{d(d-1)\over 2 L^2}\,\,,
\label{Lambda_AdS_p+1}
\eeq
which is negative. It follows from (\ref{Ricci_Cosm_Constant}) and (\ref{Lambda_AdS_p+1}) that the scalar curvature for the $AdS_{d+1}$ space with radius $L$  is given by:
\beq
R\,=\,-{d(d+1)\over L^2}\,\,.
\eeq
In these lectures we will be mostly interested in gauge theories in $3+1$ dimensions, which corresponds to taking $d=4$ in our formulas. The dual geometry found above for this case is $AdS_5$. This was precisely the system studied  in \cite{Maldacena:1997re} by Maldacena, who conjectured that the dual QFT is super Yang-Mills theory with four supersymmetries  (${\cal N}=4$ SYM).

\subsection{Counting the degrees of freedom in AdS}
\label{subsec:2.1}
After having identified the $AdS$ space as the gravity dual of a field theory with conformal invariance, we can refine the argument of section \ref{sec:1} to match the number of  degrees of freedom of both sides of the duality. 

Let us consider first the QFT side. To regulate the theory we put both a UV and IR regulator. We place the system in a spatial box of size $R$ (which serves as an IR cutoff)  and we introduce a lattice spacing $\epsilon$ that acts as a UV regulator. In $d$ spacetime dimensions the system has $R^{d-1}/\epsilon^{d-1}$ cells. Let $c_{QFT}$ be the number of degrees of freedom per lattice site, which we will refer to as the central charge. Then, the total number of degrees of freedom of the QFT is:
\beq
N_{dof}^{QFT}= \Big({R\over \epsilon}\Big)^{d-1}\,\,c_{QFT}\,\,.
\eeq
The central charge is one of the main quantities that characterize a CFT. If the CFT is  a 
$SU(N)$ gauge field theory,  such as the theory with four supersymmetries which will be described below, the fields are $N\times N$ matrices in the adjoint representation which, for large $N$, contain $N^2$ independent components. Thus, in  these $SU(N)$ CFT's the central charge scales as  $c_{SU(N)}\sim N^2$.

Let us now compute the number of degrees of freedom of the $AdS_{d+1}$ solution. According to the holographic principle and to the Bekenstein-Hawking formula (\ref{BH-formula}), the number of degrees of freedom contained in a certain region is equal to  the maximum entropy, given by
\beq
N_{dof}^{AdS}={A_{\partial}\over 4 G_N}\,\,,
\eeq
with $A_{\partial}$ being the area of the region at boundary $z\to 0$ of  $AdS_{d+1}$. Let us evaluate $A_{\partial}$ by integrating the volume element corresponding to the metric (\ref{AdS_metric}) at a slice $z=\epsilon\to 0$:
\beq
A_{\partial}\,=\,\int_{{\mathbb R}^{d-1},\, z=\epsilon}\,
 d^{d-1}\,x\,\sqrt{g}\,=\,\Big({L\over \epsilon}\Big)^{d-1}\,\
 \int_{{\mathbb R}^{d-1}}\, d^{d-1}\,x\,\,.
 \label{A_delta_unregularized}
 \eeq
The integral on the right-hand-side of (\ref{A_delta_unregularized}) is the
the volume of ${\mathbb R}^{d-1}$, which  is infinite. As we did on the QFT side, we regulate it by putting the system in  a box of size $R$:
\beq
\int_{{\mathbb R}^{d-1}}\, d^{d-1}\,x=R^{d-1}\,\,.
\eeq
Thus, the area of the $A_{\partial}$ is given by:
\beq
A_{\partial}\,=\,\Big({RL\over \epsilon}\Big)^{d-1}\,\,.
\eeq
Let us next  introduce the Planck length $l_P$ and the Planck mass $M_P$ for a gravity theory in $d+1$ dimensions  as:
\beq
G_N\,=\,(l_P)^{d-1}\,=\,{1\over (M_P)^{d-1}}\,\,.
\label{Planck_l_M}
\eeq
Then, the number of degrees of freedom of the $AdS_{d+1}$ space is:
\beq
N_{dof}^{AdS}={1\over 4}\, \Big({R\over \epsilon}\Big)^{d-1}\,
 \Big({L\over l_P}\Big)^{d-1}\,\,.
\eeq
By comparing $N_{dof}^{QFT}$ and $N_{dof}^{AdS}$  we conclude that they scale in the same way with the IR and UV cutoffs  $R$ and $\epsilon$ and we can identify:
\beq
{1\over 4}\,\,\, \Big({L\over l_P}\Big)^{d-1}\,=\,  c_{QFT}\,\,.
\label{holo_centralCharge}
 \eeq
This gives the matching  condition between gravity and QFT that we were looking for. Notice that a theory is (semi)classical when the coefficient multiplying its action is large. In this case the path integral is dominated by a saddle point. The action of our gravity theory in the $AdS_{d+1}$ space of radius $L$ contains a factor $ L^{d-1}/G_N$. 
Thus, taking into account the definition of the Planck length in (\ref{Planck_l_M}), we conclude that the classical  gravity theory is reliable if:
\beq
{\rm classical\,\, gravity\,\, in\,\, AdS}\to \Big({L\over l_P}\Big)^{d-1} \gg 1\,\,,
\eeq
which happens when the $AdS$ radius is large in Planck units. Since the scalar curvature 
goes like $1/L^2$, the curvature is small in Planck units. Thus, a QFT has a classical gravity dual when $c_{QFT}$ is large, or equivalently if there is a large number of degrees of freedom per unit volume or a large number of species (which corresponds to large $N$ for $SU(N)$ gauge theories). 

\section{String theory basics}

As mentioned in the introduction, the AdS/CFT correspondence was originally discovered in the context of string theory. Although it can be formulated without making reference to its stringy origin, it is however very convenient to know its connection to the theory of relativistic strings in order to acquire a deep understanding of the duality. We will start reviewing qualitatively this connection in this section (refs \cite{GSW, Polchinski,BBM, Elias, Zwiebach,Ibanez_Uranga}  contain systematic expositions of string theory with different levels of technical details). 

\begin{figure}[ht]
\center
\includegraphics[width=0.4\textwidth]{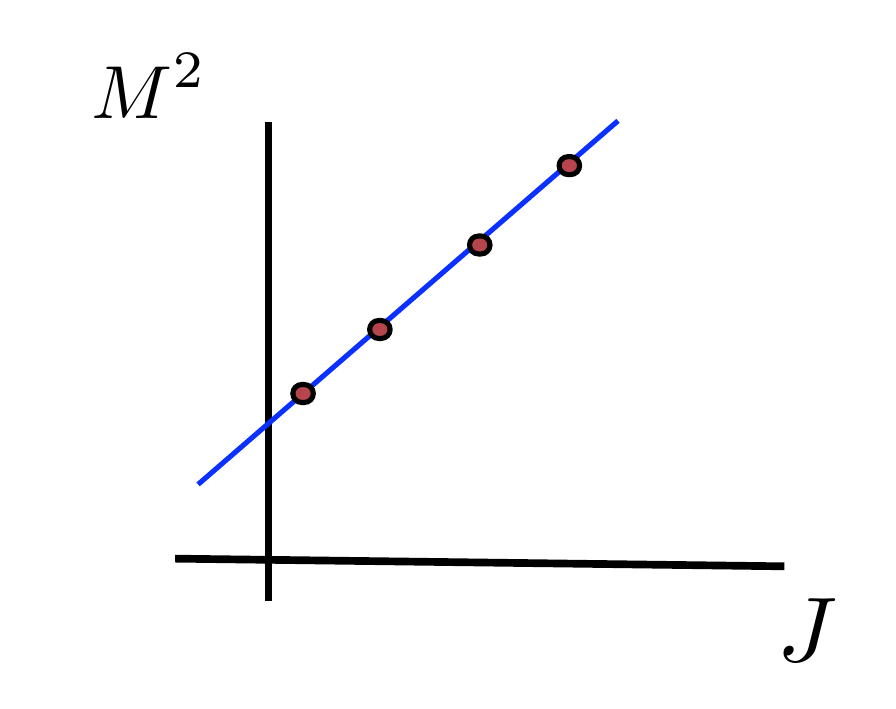}
\caption{In a Regge trajectory the mass square $M^2$ of the particles grows linearly with their spin $J$. } 
\label{Regge-trajectory}
\end{figure}

Historically, string theory was introduced in the sixties as an attempt to describe the hadronic resonances of high spin observed in the experiments. Experimentally, the mass square of these particles is linearly related to its spin $J$:
\beq
M^2\sim J\,\,.
\label{Regge_M2-J}
\eeq
It is  then said that the hadrons are distributed along Regge trajectories (see figure \ref{Regge-trajectory}). String theory was introduced to reproduce this behavior. Actually, it is not difficult to verify qualitatively that the rotational degree of freedom of the relativistic string gives rise to Regge trajectories  as in (\ref{Regge_M2-J}). Indeed, let us suppose that we have an open string with length $L$ and tension $T$ which is rotating around its center of mass. The mass of this object would be 
$M\sim T\,L$, whereas its angular momentum $J$ would be $J\sim P\,L$, with $P$ being its linear momentum. In a relativistic theory $P\sim M$, which implies that 
$J\sim P\,L\,\sim M\,L\sim T^{-1}\,M^2$ or, equivalently, $M^2\sim T\,J$. Thus, we reproduce the Regge behavior (\ref{Regge_M2-J}) with the slope being proportional to the string tension $T$. 

The basic object of string theory is an object extended along some characteristic distance $l_s$. Therefore, the  theory is non-local. It becomes local in the point-like  limit in which the size $l_s\to 0$.  The rotation degree of freedom of the string gives rise to Regge trajectories similar to those observed experimentally. In modern language one can regard a meson as a quark-antiquark  pair joined by a string. The energy of such a configuration grows linearly with the length, and this constitutes a model of confinement.

The classical description of a relativistic string is directly inspired from that of a point particle in the special theory of relativity. Indeed, let us consider a relativistic point particle  of mass $m$ moving in a flat spacetime with Minkowski metric $\eta_{\mu\nu}$. As it moves the particle describes a curve in spacetime (the so-called worldline), which can be represented by a function of the type:
\beq
x^{\mu}\,=\,x^{\mu}(\tau)\,\,,
\eeq
where $x^{\mu}$ is the coordinate in the space in which the point particle is moving (the target space) and  $\tau$  parameterizes the path of the particle (the worldline coordinate). The action  of the particle is proportional to the integral of the  line element along the trajectory in spacetime, with the coefficient being given by the mass $m$ of the particle:
\beq
S\,=\,-m\,\int ds\,=\,
-m\,\int_{\tau_0}^{\tau_1}\,d\tau \,\,\sqrt{-\eta_{\mu\nu}\,\dot x^{\mu}\,\dot x^{\nu}}\,\,.
\label{NG-particle}
\eeq

\begin{figure}[ht]
\center
\qquad\qquad\includegraphics[width=0.4\textwidth]{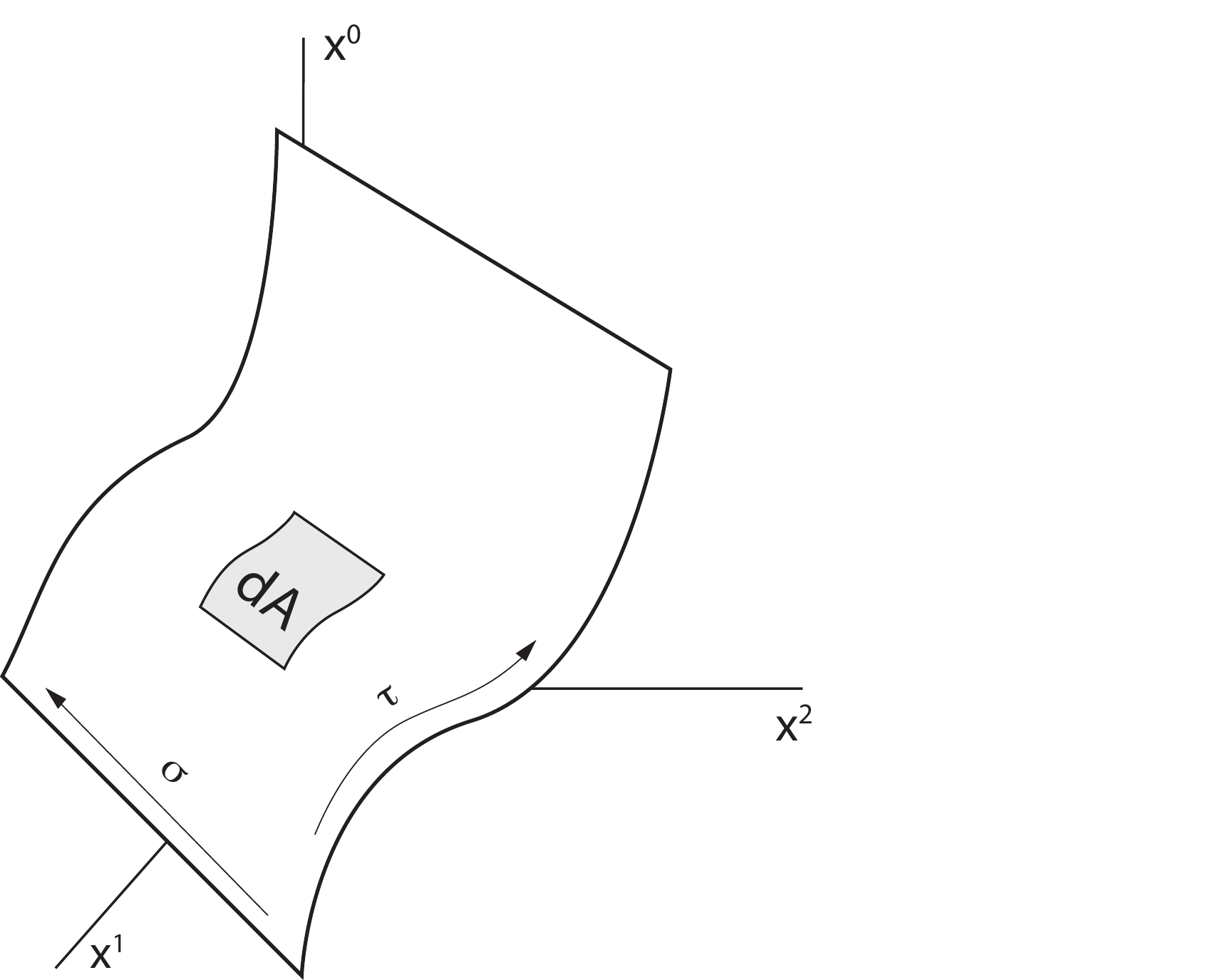}
\caption{The string sweeps in spacetime a two-dimensional worldsheet. } 
\label{worldsheet}
\end{figure}

Let us now consider a relativistic string. A one-dimensional object moving in spacetime describes a surface (the so-called worldsheet). Let $dA$ be the area element of the worldsheet (see figure \ref{worldsheet}). Then, the analogue of the action (\ref{NG-particle}) for  a string is the so-called Nambu-Goto action:
\beq
S_{NG}\,=\,-T\,\int \,dA\,\,,
\label{NG-string}
\eeq
where $T$ is the tension of the string, given by:
\beq
T\,=\,{1\over 2\pi\alpha'}\,\,,
\eeq
and $\alpha'$ is the Regge slope ($\alpha'$ has units of $(length)^2 $ or $(mass)^{-2}$). The string length and mass are defined as:
\beq
l_s\,=\,\sqrt{\alpha'}\,=\,{1\over M_s}\,\,.
\label{ls}
\eeq

Then, the string tension is related to $l_s$ and $M_s$ as:
\beq
T\,=\,{1\over 2\pi l_s^2}\,=\,{M_s^2\over 2\pi}\,\,.
\eeq
Let us write more explicitly the Nambu-Goto action (\ref{NG-string}). With this purpose we will take two coordinates $\xi^{\alpha}$ ($\alpha=0,1$) to parameterize the worldsheet 
$\Sigma$ ( $(\xi^0, \xi^1)\,=\,(\tau, \sigma)$). Let us assume that the string moves in a target space  ${\cal M}$ with metric $G_{\mu\nu}$. The embedding of $\Sigma$ in the spacetime ${\cal M}$  is characterized by a map $\Sigma\rightarrow {\cal M}$ with $\xi^{\alpha}\rightarrow X^{\mu}(\xi^{\alpha})$. The induced metric on $\Sigma$ is:
\beq
\hat G_{\alpha\beta}\,\equiv \,G_{\mu\nu}\,\partial_{\alpha}\, X^{\mu}\,
\partial_{\beta}\, X^{\nu}\,\,.
\eeq
Thus, the Nambu-Goto action of the string  is:
\beq
S_{NG}\,=\,-T\,\int \sqrt{-\det \hat G_{\alpha\beta}}\,\,\,d^2\xi\,\,.
\label{NG-coordinates}
\eeq
The action  (\ref{NG-coordinates}) depends non-linearly on the embedding functions $X^{\mu} (\tau, \sigma)$. The classical equations of motion derived from (\ref{NG-coordinates}) are partial differential equations which, remarkably, can be solved for a flat target spacetime with $G_{\mu\nu}=\eta_{\mu\nu}$, both for Neumann  and Dirichlet  boundary conditions. Actually, the functions $X^{\mu} (\tau, \sigma)$ can be represented in a Fourier expansion as an infinite superposition of oscillation modes, much as in the string of a violin.

The quantization of the string can be carried out by using the standard methods in quantum physics. The simplest way is just by canonical quantization, \ie\ by considering that the $X^{\mu}$ are operators and by imposing canonical commutation relations between coordinates and momenta. As a result one finds that the different oscillation modes can be interpreted as particles and that the spectrum of the string contains an infinite tower of particles with growing masses and spins that are organized in Regge trajectories, with $1/ l_s$ being the mass gap.

The close scrutiny of the consistency conditions of the quantum string reveals many surprises (quantizing the string is like opening Pandora's box!). First of all, the mass spectrum contains tachyons (particles with $m^2<0$), which is a signal of instability. In order to avoid this problem one must consider a string which has also fermionic coordinates and require that the system is supersymmetric (\ie\ that there is a symmetry between bosonic and fermionic degrees of freedom). This generalization of the bosonic string (\ref{NG-coordinates}) is the so-called superstring. Another consistency requirement imposed by the quantization is that the number $D$ of dimensions of the space in which the string is moving is fixed. For a superstring $D=10$. This does not mean that the extra dimensions have the same meaning as the ordinary ones of the four-dimensional Minkowski spacetime. Actually,  the extra dimensions should be regarded as defining a configuration space (as  the phase space in classical mechanics does). We will see below that this is precisely the interpretation that they have in the context of the AdS/CFT correspondence. 

Another important piece of information about the nature of string theory is obtained from the analysis of the spectrum of massless particles. Massive particles have a mass which a multiple of $1/l_s$ and, therefore, they become unobservable in the low-energy limit $l_s\to 0$. Then, after eliminating the tachyons by using supersymmetry, the massless particles are the low-lying excitations of the spectrum. We must distinguish the case of open and closed strings. The spectrum of open strings contains massless particles of spin one with the couplings needed to have gauge symmetry. These particles can be naturally identified with gauge bosons (photons, gluons,...). The big surprise comes when looking at the spectrum of  closed strings, since  it contains a particle of spin two and zero mass which can only be interpreted as the graviton (the quantum of gravity). Besides, quantum consistency of the propagation of the string in a curved space implies Einstein equations in ten dimension plus corrections:
\beq
R_{\mu\nu}+\cdots=0\,\,.
\eeq
Then, one is led to conclude that string theory is not a theory of hadrons but a  theory of quantum gravity!!.  Thus, the string length $l_s$ should be of the order of the Planck length $l_P$ and not of the order of the hadronic scale $\sim$ 1 fm).  Elementary strings without internal structure and zero thickness were born for the wrong purpose. Moreover, in addition to the graviton, the massless spectrum contains antisymmetric tensor fields of the form $A_{\mu_1\cdots\mu_{p+1}}$, which will become very relevant in formulating the AdS/CFT correspondence. 

\begin{figure}[ht]
\center
\includegraphics[width=0.6\textwidth]{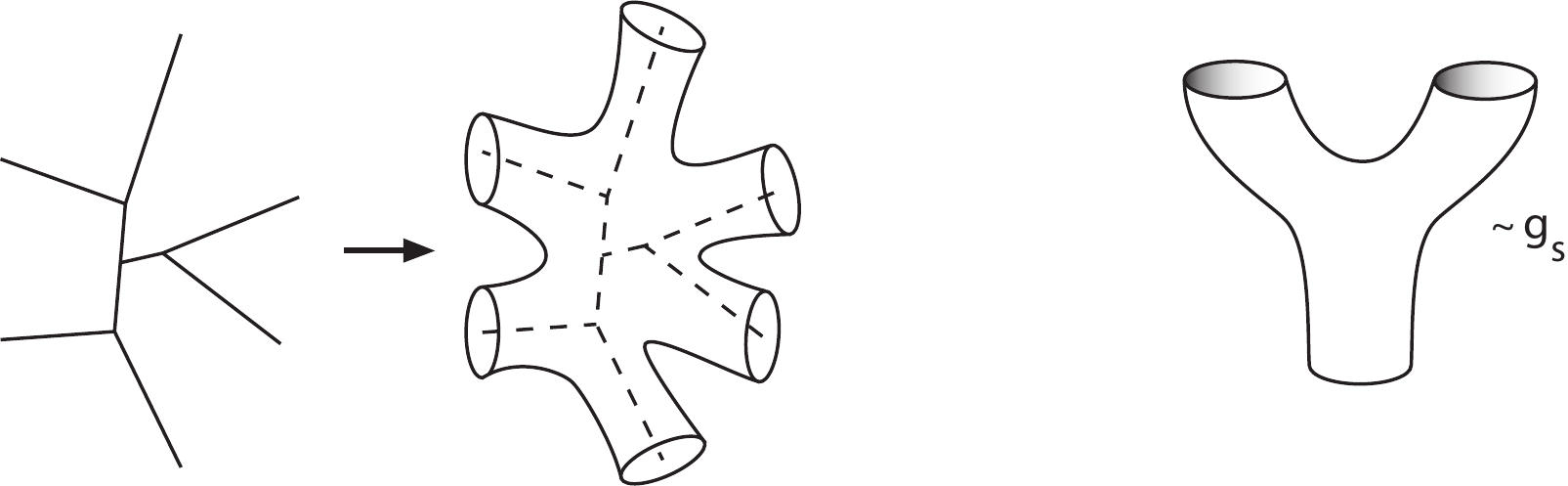}
\caption{A vertex in field theory corresponds to a two-dimensional surface with boundaries (left). The triple vertex for three closed strings is represented  on the right as a pants surface. } 
\label{thick_and_triple_vertex}
\end{figure}

\begin{figure}[ht]
\center
\includegraphics[width=0.9\textwidth]{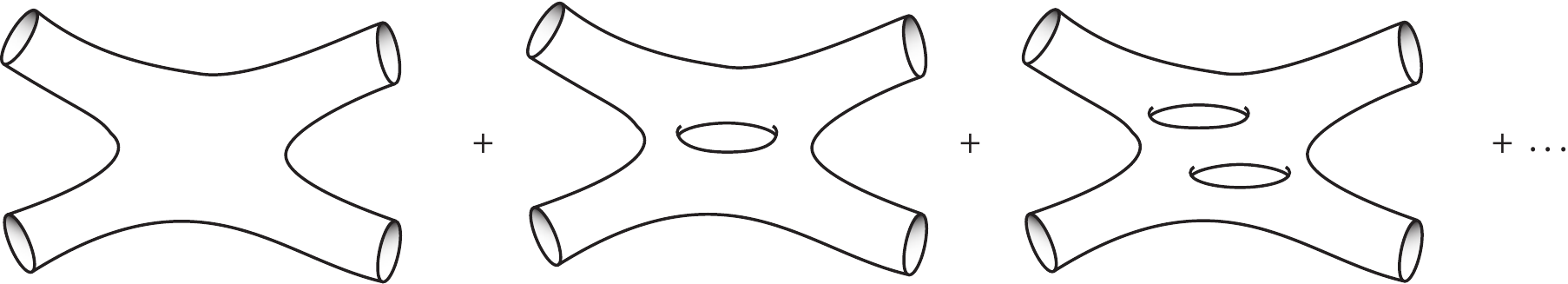}
\caption{Perturbative expansion of the amplitude corresponding to four closed strings. } 
\label{stringperturbation}
\end{figure}

As strings propagate through the target space ${\cal M}$, they can suffer interactions by splitting in two or more parts or by joining with other strings. The worldsheet for one of these interactions is just a two-dimensional surface with holes and boundaries which can be thought as a vertex in QFT perturbation theory in which the lines has been thickened, as shown in figure \ref{thick_and_triple_vertex}. In closed string theory the basic interaction is a string splitting into two (or the inverse process of two closed strings merging into one). The corresponding worldsheet has the form of pants, which can be obtained by thickening a triple vertex in QFT. A loop in string perturbation theory can be obtained with two triple vertices, which produces a worldsheet which is a Riemann surface with a hole. Higher order terms in perturbation theory correspond to having more holes on the worldsheet (see figure \ref{stringperturbation}). Therefore, the perturbative series in string theory is a topological expansion!. In the topology of two-dimensional  surfaces, the number $h$ of holes (or handles) of the surface  is called the genus of the surface. In string perturbation theory, $h$ is just the number of string loops. 

Let us suppose that we weigh every triple vertex with a string coupling constant  coupling $g_s$.  It is clear from the discussion above that  string perturbation theory is an expansion of the type:
\beq
{\cal A}\,=\,\sum_{h=0}^{\infty}\, g_s^{2h-2}\,F_h(\alpha')\,\,,
\label{string_pert_series}
\eeq
where ${\cal A}$ is some amplitude. In the next section we will identify a topological expansion in gauge theories which has the same structure as the series (\ref{string_pert_series}). 

\section{Large $N$ expansion in gauge theories}
Let us consider $U(N)$  Yang-Mills theory with lagrangian
\beq
{\cal L}\,=\,-{1\over g^2}\,{\rm Tr}\big[ F_{\mu\nu} F^{\mu\nu}\big]\,\,,
\eeq
where the non-abelian gauge field strength is given by:
\beq
F_{\mu\nu}\,=\,\partial_{\mu}\,A_{\nu}-\partial_{\nu} A_{\mu}\,+\,
[A_{\mu}, A_{\nu}]\,\,.
\eeq
In the equations above $A_{\mu}$ is an $N\times N$ matrix whose elements can be written as $A_{\mu\,,\,b}^{a}$, where $a,b$ are indices that run from $1$ to $N$. Let us rewrite ${\cal L}$ as:
\beq
{\cal L}\,=\,-{N\over \lambda}\,{\rm Tr}\big[ F_{\mu\nu} F^{\mu\nu}\big]\,\,,
\eeq
where $\lambda=g^2 N$ is the so-called 't Hooft coupling. The  't Hooft  expansion \cite{tHooft:1973jz} corresponds to keeping $\lambda$ fixed and to perform an expansion of the amplitudes in powers of $N$. It turns out that the different powers of $N$ correspond to the different topologies. One can prove this by
adopting a double line notation for the gauge propagator. Then, one can verify that 
the color lines form the perimeter of an oriented polygon (a face). 
Polygons join at  a common edges in such a way that every vacuum 
graph is associated to a  triangulated two-dimensional surface formed by sewing all polygons along the edges. 

\begin{figure}[ht]
\center
\includegraphics[width=0.43\textwidth]{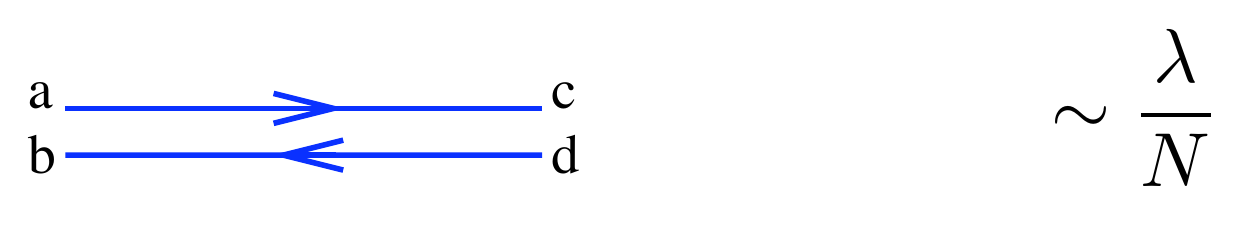}
\qquad\qquad
\includegraphics[width=0.63\textwidth]{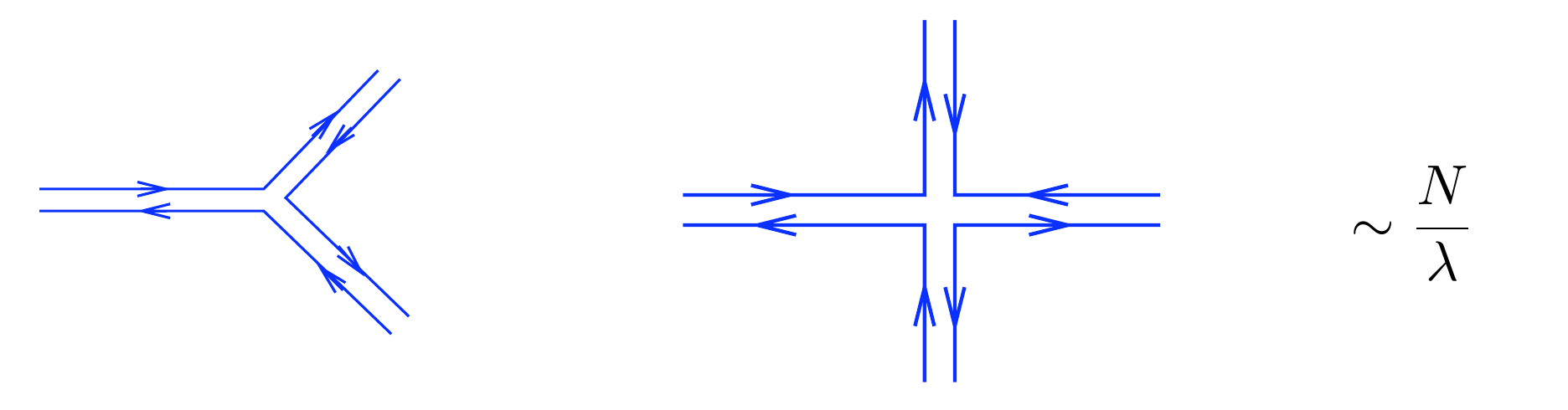}
\caption{Gauge propagator and vertices in the double line notation of Yang-Mills  $U(N)$ theories. } 
\label{largeNYM.pdf}
\end{figure}

\begin{figure}[ht]
\center
\includegraphics[width=0.43\textwidth]{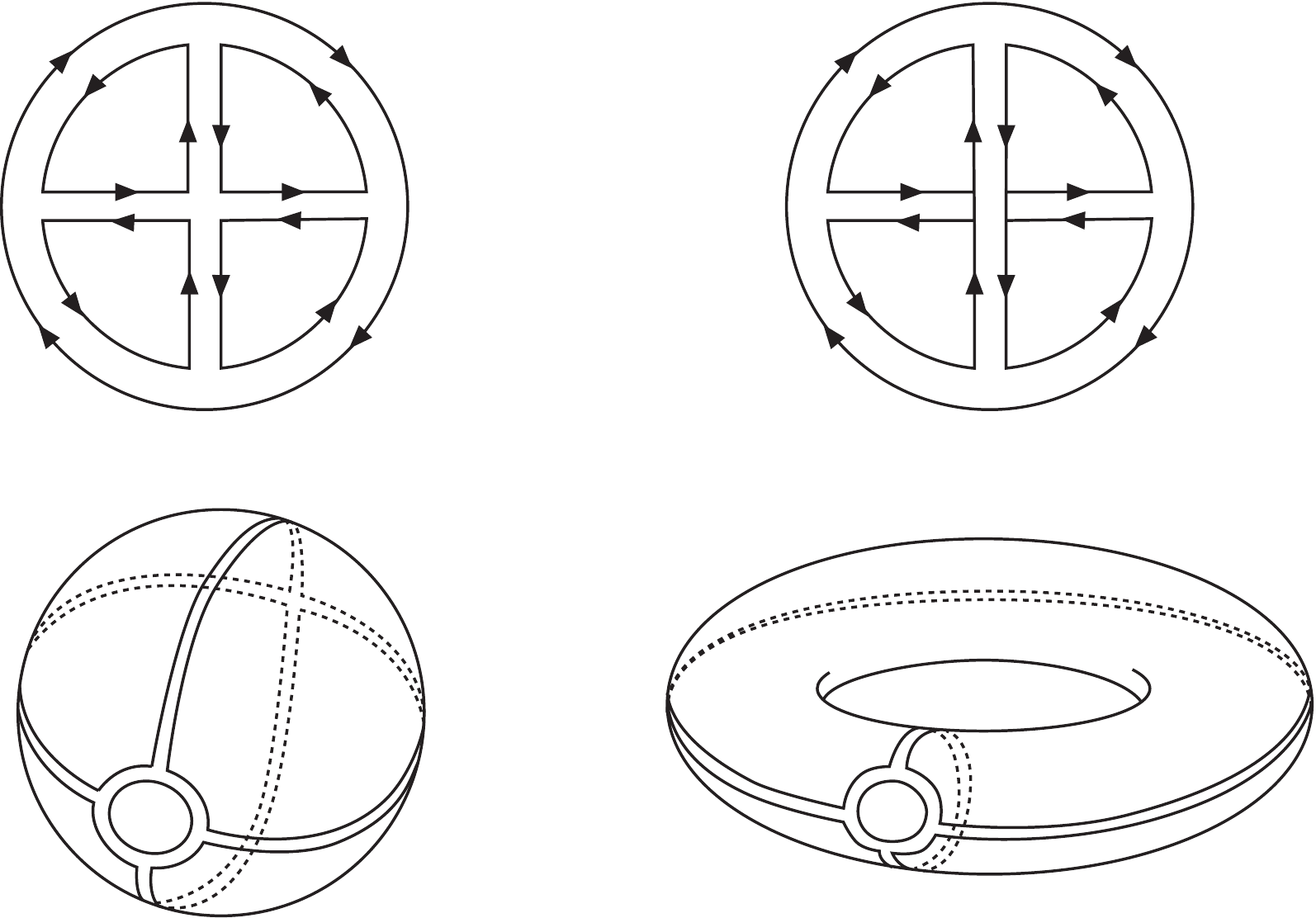}
\caption{The planar diagram  on the left can be drawn on a sphere, whereas the non-planar diagram on the right must be drawn on a torus.} 
\label{planar_and_nonplanar}
\end{figure}

It is not difficult to find the powers of $N$ and $\lambda$ appearing in a given diagram $D$ with no external lines. The contributions of the gauge propagator and the vertices are displayed in figure \ref{largeNYM.pdf}.  Moreover, every index loop contributes with a power of $N$.  Suppose that $E$ is the number of propagators (edges) of $D$ connecting two vertices,  $V$  is the number of vertices  and $F$  is the number of index loops (faces). Then, one can show that:
\beq
D\,\sim \,\Big({\lambda\over N}\Big)^{E}\,\Big({N\over \lambda}\Big)^{V}\,N^{F}\,=\,
N^{F-E+V}\,\lambda^{E-V}\,\,.
\label{D-largeN}
\eeq
Let us denote by $\chi$ the combination of $F$, $E$ and $V$ that appears in the exponent of $N$ in (\ref{D-largeN}):
\beq
F-E+V=\chi\,\,,
\eeq
which is nothing but the Euler characteristic of the triangulation, which only depends on the topology of the surface. If the diagram triangulates a surface with $h$ handles (an $h$-torus), $\chi$ is given by:
\beq
\chi=2-2h\,\,.
\eeq
Thus the diagrams are weighed  with a power of $N$ determined by the number of handles of the surface that they triangulate (see figure \ref{planar_and_nonplanar}). The planar diagrams are those that can be drawn on a piece of paper without self-crossing. They correspond to $h=0$ and their contribution is of the type $N^2\lambda^n$, for some power $n$ which depends on the diagram considered. Clearly, as the dependence on $N$ goes like $N^{2-2h}$ the diagrams with $h=0$ are the dominant ones in the large $N$ expansion. For this reason the large $N$ limit is also called the planar limit of the gauge theory (see \cite{Manohar:1998xv} for more details and examples).

The effective action $Z$ can be obtained by computing the sum over the connected vacuum-to-vacuum diagrams. It has the structure:
\beq
\log Z\,=\sum_{h=0}^{\infty}\,N^{2-2h}\,
\sum_{l=0}^{\infty}\,c_{l,h}\,\lambda^{l}\,=\,
\sum_{h=0}^{\infty}\,N^{2-2h}\,f_h(\lambda)\,\,,
\eeq
where $f_h(\lambda)$ is the sum of Feynman diagrams that can be drawn in a surface of genus $h$. This clearly suggest a connection with the perturbative expansion of  string theory  written in (\ref{string_pert_series}) if one identifies the string coupling constant $g_s$ as:
\beq
g_s\sim {1\over N}\,\,.
\eeq
Heuristically one can say that gauge theory diagrams triangulate the worldsheet of an effective string. 
The AdS/CFT correspondence is a concrete realization of this connection in the limit $(N, \lambda)\to\infty$ (\ie\ for planar theories in the strongly coupled regime). This dual stringy description of gauge theory is nothing but a quantum version of the familiar description of electromagnetism in terms of the string-like lines of force. 

\section{D-branes}

\begin{figure}[ht]
\center
\includegraphics[width=0.25\textwidth]{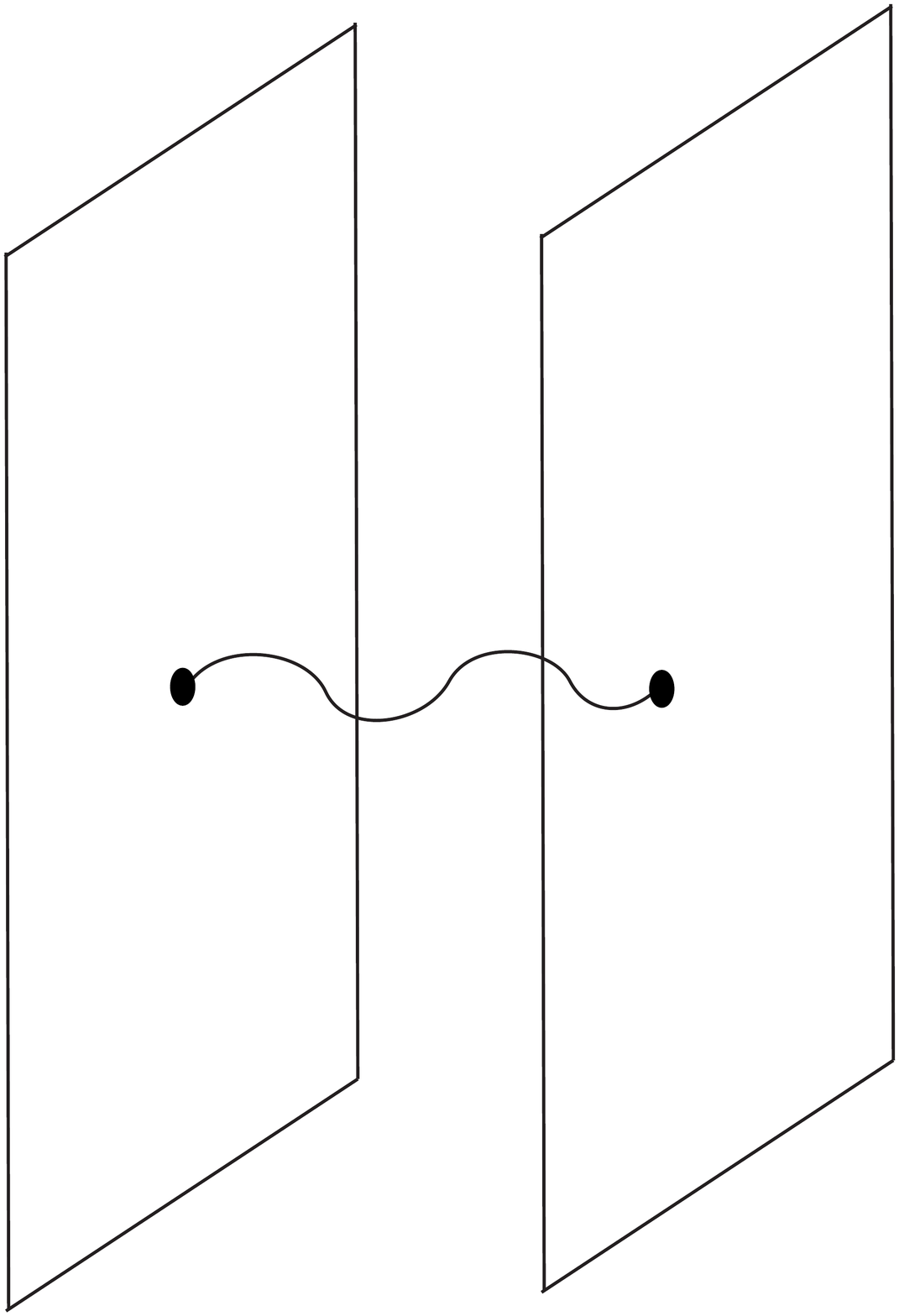}
\qquad\qquad\qquad
\includegraphics[width=0.35\textwidth]{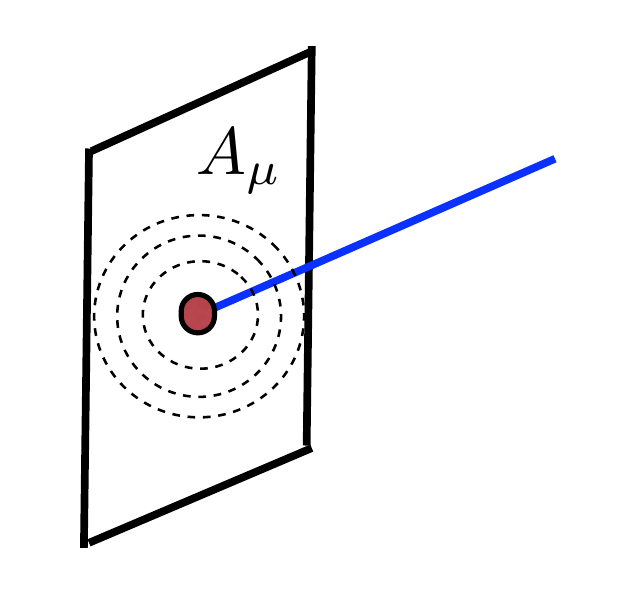}
\caption{The D-branes are hyperplanes where open strings can end. The endpoints of the strings source gauge fields $A_{\mu}$ on the D-brane worldvolume.} 
\label{Dbranes}
\end{figure}

Besides the perturbative structure reviewed above, string theories have  a non-perturbative sector, which plays a crucial role in connecting them with gauge theories. The  relevant objects  in this non-perturbative sector are the solitons, which  are extended objects. The most important solitons for the AdS/CFT correspondence   are the  Dp-branes, that are objects extended in $p+1$ directions  ($p$ spatial + time). The Dp-branes can be defined as hypersurfaces where strings end (see figure \ref{Dbranes}). They can be obtained by quantizing the string with fixed ends along hyperplanes (Dirichlet boundary conditions). In addition, they can also be understood as objects charged under the antisymmetric tensor fields $A_{\mu_1\cdots\mu_{p+1}}$ of string theory, which naturally couple to the Dp-brane worldvolume as:
\beq
A_{\mu_1\cdots\mu_{p+1}}\to
\int_{{\cal M}_{p+1}}\,A_{\mu_1\cdots\mu_{p+1}}\,\, dx^{\mu_1}\cdots  dx^{\mu_{p+1}}\,\,,
\eeq
with ${\cal M}_{p+1}$  being the worldvolume of the Dp-brane. The Dp-branes are dynamical objects
that can move and get excited. Schematically, their action takes the form:
\beq
S_{Dp}\,=\,-T_{Dp}\,\int d^{p+1} x\,[\cdots]\,\,,
\eeq
with $T_{Dp}$ being the tension of the Dp-brane, which in terms of the string coupling constant $g_s$ and the  string  length $l_s$ is given by:
\beq
T_{Dp}={1\over (2\pi)^p \,g_s \, l_s^{p+1}}\,\,.
\label{TDp}
\eeq

Notice from (\ref{TDp}) that $T_{Dp}\sim g_s^{-1}$,  which confirms that the Dp-branes are non-perturbative objects in string theory. The dependence of $T_{Dp}$ on $l_s$ in (\ref{TDp}) is fixed by dimensional analysis. 

The Dp-branes can have two classes of excitations. The first one correspond to  rigid motions and deformations of their shape. These degrees of freedom can be parameterized by the  $9-p$  coordinates $\phi^{i}\,\,\,(i=1,\cdots,9-p)$ transverse to the ($p+1$)-dimensional worldvolume in the ten-dimensional target space. The $\phi^{i}$'s are just  scalar fields on the  Dp-brane worldvolume. Besides, the Dp-branes can have internal excitations. To get a clue on how to represent these excitations, let us recall that the endpoint of the string is a charge. When  there is a charge,  a gauge field is sourced, as illustrated in figure \ref{Dbranes}. Then, it follows that a Dp-brane has an abelian gauge field $A_{\mu}$ $(\mu=0,\cdots, p)$ living  in its worldvolume. The action of the Dp-brane which takes into account these two types of excitations is the so-called Dirac-Born-Infeld action, which can be written as:
\beq
S_{DBI}\,=\,-T_{Dp}\,\int d^{p+1} x \sqrt{-\det (g_{\mu\nu}+2\pi\, l_s^2 \,F_{\mu\nu})}\,\,,
\label{DBI_action}
\eeq
where $g_{\mu\nu}$ is the induced metric on the worldvolume and 
$F_{\mu\nu}=\partial_{\mu}\,A_{\nu}\,-\,\partial_{\nu}\,A_{\mu}$ is the field strength of the gauge field $A_{\mu}$. When the worldvolume gauge field is not excited ($F_{\mu\nu}=0$) the DBI action (\ref{DBI_action}) becomes the natural generalization of the Nambu-Goto action (\ref{NG-coordinates}) for an object extended along $p$ spatial directions.

Let us consider  a Dp-brane in flat space. The induced metric in this case takes the form:
\beq
g_{\mu\nu}\,=\,\eta_{\mu\nu}+(2\pi l_s^2)^2\,\partial_{\mu}\phi^i\,\partial_{\nu}\phi^i\,\,,
\eeq
where the $\phi^i$ are the coordinates that parameterize the embedding of the brane. 
Let us now expand the square root of (\ref{DBI_action}) in powers of $F_{\mu\nu}$ and $\partial_\mu\phi$.  The quadratic terms in this expansion can be written as:
\beq
S_{DBI}^{(2)}\,=\,-{1\over g^2_{YM}}\,\,
\Big(\,{1\over 4}\,F_{\mu\nu}F^{\mu\nu}\,+\,{1\over 2}\,
\partial_{\mu}\phi^i\,\partial^{\mu}\phi^i\,+\,\cdots\,\Big)\,\,,
\label{DBI-2nd_order}
\eeq
which is just the ordinary action of a gauge field and $9-p$ scalar fields. In (\ref{DBI-2nd_order})
 $g_{YM}$ is the Yang-Mills coupling, which, in terms of $l_s$ and $g_s$, is given by:
\beq
g^2_{YM}\,=\,2(2\pi)^{p-2}\,l_s^{p-3}\,g_s\,\,.
\label{gYM-p}
\eeq
In addition to $A_{\mu}$ and $\phi$, in superstring theory the branes have fermionic excitations, which can be represented in terms of fermionic fields.

We have just discovered one of the most important features of the Dp-branes: they contain a gauge theory living in their wordlvolume!. In the case of a single Dp-brane the gauge group is $U(1)$, as in (\ref{DBI-2nd_order}).  However, if we consider a stack of coincident parallel Dp-branes the gauge group gets promoted to $U(N)$. Indeed, one can verify in this case that the fields $A_{\mu}$ and $\phi^i$ are matrices transforming in the adjoint representation of $U(N)$. The non-diagonal elements of these fields correspond to excitations connecting different branes, whereas the diagonal ones are excitations on a single brane. Actually, the $U(1)$ component of the $U(N)$ gauge theory can be decoupled from the $SU(N)$ fields. Therefore, we conclude that a stack of  $N$ Dp-branes realizes a $SU(N)$ gauge theory in $p+1$ dimensions.

The D-branes have brought a completely new perspective on gauge theories which gives rise to what is called brane engineering. The geometric realization of the gauge symmetry that they  provide allows to perform a series of transformations which lead to discover new unexpected  properties of gauge theories. For example, 
one can move the branes, place them in different spaces, etc. In the gauge theory, these transformations lead to dualities, reductions of the amount of supersymmetry, changes of  the field content, etc. In general we have a novel geometric insight on gauge dynamics. The AdS/CFT correspondence is an important outcome of this new realization of the gauge symmetry.

The particular case of the D3-branes is specially relevant in what follows. In this case we have a $3+1$ worldvolume and $10-4=6$ scalar fields. The corresponding four-dimensional $SU(N)$ gauge theory can be identified with super Yang-Mills theory with four supersymmetries (${\cal N}=4$ SYM). This theory is an exact CFT at the quantum level  and will be the basic example of the AdS/CFT correspondence. Notice from (\ref{gYM-p}) that the Yang-Mills coupling is dimensionless in this case and is related to the string coupling constant $g_s$ as:
\beq
g^2_{YM}\,=\,4\pi g_s\,\,,
\qquad\qquad
({\rm D3-branes})\,\,.
\label{gYM-p3}
\eeq

\section{D-branes and gravity}

\begin{figure}[ht]
\center
\includegraphics[width=0.9\textwidth]{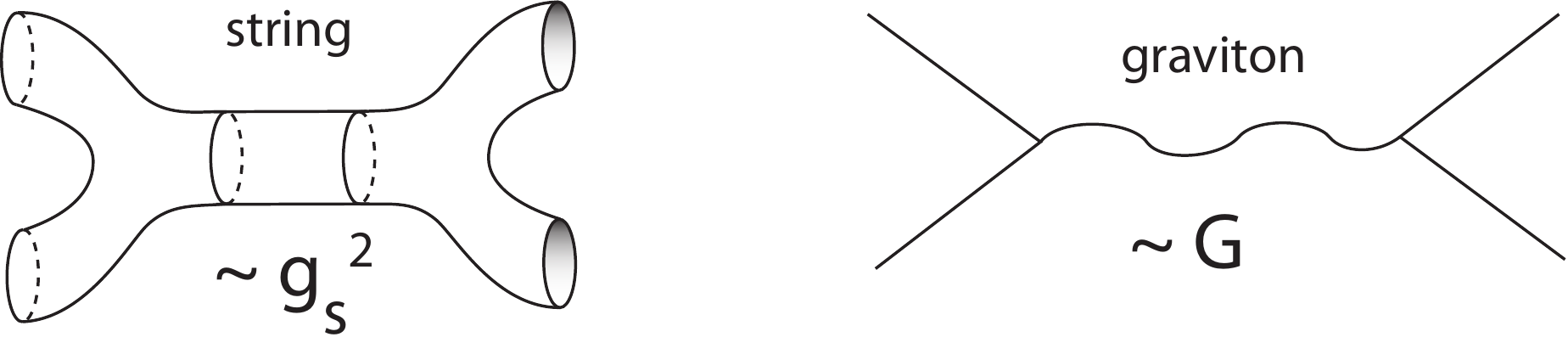}
\caption{In string theory the exchange of a graviton is obtained as the low-energy limit of the exchange of a closed string. } 
\label{Newton_string.pdf}
\end{figure}

String theory is a gravity theory in which all types of matter distort the spacetime. This distortion is determined by  solving Einstein equations which follow from the action:
\beq
S\,=\,{1\over 16\pi\, G}\,\int d^{10} x\,\sqrt{-g}\, R\,+\,\cdots\,\,.
\label{Einstein-Hilbert}
\eeq
The ten-dimensional Newton constant  in (\ref{Einstein-Hilbert}) is related to string parameters as:
\beq
16\pi\, G\,=\,(2\pi)^7\,g_s^2\,l_s^8\,\,.
\label{10d_Newton_constant}
\eeq
The dependence of $G$ on $l_s$ in (\ref{10d_Newton_constant})  follows from dimensional analysis. The dependence on $g_s$ follows by comparing two-to-two scattering amplitudes in string theory and in supergravity (see figure \ref{Newton_string.pdf}). 

The Dp-branes (and other extended objects) are solutions of the Einstein equations. Let us consider these solutions at the linearized level of weak gravity (\ie\ when we are far from the object). The linearized metric for a point-like object in a $D$-dimensional  spacetime is:
\beq
ds^2\,\approx\,-(1+2\varphi)\,dt^2\,+\,\Big(1-{2\over D-3}\varphi\,\Big)\,
\Big(dx_1^2\,+\,\cdots\,+\, dx_{D-1}^2\Big)\,\,,
\label{point_like_metric}
\eeq
where the function  $\varphi$ parameterizes the deviation of the metric from the flat Minkowski metric in $D$ dimensions. The specific form (\ref{point_like_metric}) can be obtained by solving the linearized Einstein equations and generalizes the standard result in general relativity  in four dimensions. As in this latter particular case, one can identify the function $\varphi$ with
the Newtonian gravitational potential, as seen by comparing the motion along a geodesic and Newton's law. Thus:
\beq
\varphi\sim {GM\over r^{D-3}}\,\,,
\eeq
where $M$ is  the mass of the particle and  $r=\sqrt{x_1^2\,+\,\cdots\,+\, x_{D-1}^2}$  is the radial coordinate of the space. The function  $\varphi$ is a solution of the Poisson equation in $D-1$ dimensions. Notice that the power $D-3$ is equal to the number $d_T$ of dimensions transverse to the object (i. e.   $d_T=D-1$) minus 2:
\beq
D-3=d_T-2\,\,.
\eeq
For an extended object (along $p$ spatial directions)  the corresponding linearized metric takes the form:
\beq
ds^2\approx(1+2\varphi)\,\big(-dt^2+dx_1^2+\cdots+ dx_{p}^2\big)+
\big(1-{2(p+1)\over D-p-3}\varphi\,\big)
\big(dx_{p+1}^2+\cdots+ dx_{D-1}^2\big)\,\,.
\label{extended_metric}
\eeq
Since now the number of transverse directions is $d_T=D-1-p$, the function  $\varphi$ must be:
\beq
\varphi\sim {GM\over r^{D-p-3}}\,\,.
\label{extended_potential}
\eeq
The  D3-brane solution is a solution of 10d supergravity corresponding to a stack of $N$ coincident D3-branes. The (exact) metric takes the form:
\beq
ds^2\,=\,H^{-{1\over 2}}\,
\Big(-dt^2+dx_1^2\,+dx_2^2\,+\, dx_{3}^2\Big)+
H^{{1\over 2}}\,(dr^2+r^2 d\Omega_5^2\,)\,\,,
\label{D3-brane-metric}
\eeq
with $d\Omega_5^2$ is the  line element of a unit ${\mathbb S}^5$. Notice that   $dr^2+r^2 d\Omega_5^2$ is the  flat metric of ${\mathbb R}^6$.  The function $H(r)$  is the so-called warp factor:
\beq
H=1+{L^4\over r^4}\,\,,
\eeq
with the constant $L$ being given by:
\beq
L^4= 4\pi g_s N l_s^4\,\,.
\label{D3-brane-radius}
\eeq
At linearized level (when $r\to\infty$) the solution (\ref{D3-brane-metric})
corresponds to the general solution in (\ref{extended_metric}) with $D=10$, $p=3$ and the function $\phi$ given by: 
\beq
\varphi=-{1\over 4}\,{L^4\over r^4}\,\,.
\eeq
Comparing this expression of $\varphi$ with the one written in (\ref{extended_potential}) for $D=10$ and  $p=3$ we conclude that:
\beq
GM\sim L^4\,\,.
\eeq
Let us find  this last relation from a different reasoning. 
As  $G\sim g_s^2\,l_s^8$ (see (\ref{10d_Newton_constant})) and  since $M$ should be the mass of a stack of $N$ D3-branes with tension $T_{D3}\sim 1/(g_s l_s^4)$, we should have:
\beq
GM\sim  g_s^2\,l_s^8 {N\over g_s l_s^4}\sim N g_s l_s^4\,\,,
\eeq
which, apart from a numerical factor is precisely the value of $L^4$ written in (\ref{D3-brane-radius}).

The geometry of this solution is asymptotically the Minkowski spacetime in 10d with a throat of infinite size (see figure \ref{Kadanoff}). In the throat we can take $r<<1$ and neglect the $1$ in the function $H$. This defines the so-called near-horizon limit. Actually, taking
\beq
H\approx {L^4\over r^4}\,\,,
\eeq
the metric (\ref{D3-brane-metric})  becomes:
\beq
ds^2={r^2\over L^2}\,(-dt^2+dx_1^2\,+dx_2^2\,+\, dx_{3}^2)+{L^2\over r^2}\,dr^2\,+\,L^2\,
d\Omega_5^2\,\,.
\label{D3-metric_near-horizon}
\eeq
To identify this near-horizon metric, let us change variable as:
\beq
r={L\over z}\,\,.
\eeq
Then, the line element (\ref{D3-metric_near-horizon}) can be rewritten as:
\beq
ds^2={L^2\over z^2}\,(-dt^2+dx_1^2\,+dx_2^2\,+\, dx_{3}^2+dz^2)
\,+\,L^2\,
d\Omega_5^2\,\,,
\eeq
which  is precisely  the metric of the product space $AdS_5\times {\mathbb S}^5$,  with $L$ being the radius of both factors.

\section{The AdS/CFT for ${\cal N}=4$ $SU(N)$ SYM}

We are now in a position to formulate the AdS/CFT correspondence, as originally  proposed by Maldacena in  \cite{Maldacena:1997re}. We have seen two different descriptions of the D3-brane: as a gauge theory and as a gravity solution.  Maldacena proposed that the two descriptions are dual to each other. More generally, it was conjectured in \cite{Maldacena:1997re} that 
${\cal N}=4$  SYM theory with gauge group $SU(N)$  is equivalent to  string theory in  $AdS_5\times S^5$. 

In order to make the duality  more precise, it is quite convenient to find the relation of the two parameters on the two sides of the correspondence. First of all, by combining (\ref{D3-brane-radius}) and (\ref{gYM-p3}) we can write the ratio between the AdS radius $L$ and the string length $l_s$ in terms of gauge theory quantities:
\beq
\Bigg({L\over l_s}\Bigg)^4 \,=\,N\,g^2_{YM}\,\,.
\label{L-ls-gYM}
\eeq

Notice that the  right-hand side of (\ref{L-ls-gYM}) is just the 't Hooft coupling $\lambda=N\,g^2_{YM}$. Therefore, we can rewrite (\ref{L-ls-gYM}) as:
\beq
{l_s^2\over L^2}\,=\,{1\over \sqrt{\lambda}}\,\,.
\label{l_s-L}
\eeq
Moreover, by combining (\ref{10d_Newton_constant}) and (\ref{gYM-p3}) we can write the ten-dimensional Newton constant $G$, and the corresponding Planck length $l_P$, in terms of $g_{YM}$, namely:
\beq
G\,=\,l_P^8\,=\, {\pi^4\over 2}\,g^4_{YM}\,l_s^8\,\,.
\eeq
Then, the relation between the Planck length and the AdS radius is given by
\beq
\Big({l_P\over L}\Big)^8\,=\,{\pi^4\over 2 N^2}\,\,.
\label{L_P-L}
\eeq
The relations (\ref{l_s-L}) and (\ref{L_P-L}) are essential to delimitate the domain of validity of the dual description in terms of {\rm classical} gravity.  As discussed above, we must require that $l_P/ L\ll 1$ if we want to avoid having quantum gravity corrections (this is equivalent to the condition of having small curvature). From (\ref{L_P-L}) is clear that one should require that $N\gg 1$. Notice this conclusion agrees with our general discussion at the end of section \ref{sec:2}. Moreover, in order to avoid stringy corrections due to the full tower of massive states of the string, we must require that
$l_s/ L\ll 1$ which, in view of (\ref{l_s-L}), means that the 't Hooft coupling must be large.  
Therefore, we conclude that the gravitational description of ${\cal N}=4$ SYM is reliable if:
\beq
N\gg 1\,\,,
\qquad\qquad
\lambda \gg 1\,\,.
\eeq
Then, the planar (large $N$), strongly coupled (large $\lambda$) SYM theory can be described as classical gravity. As a first check of the correspondence, let us look at the symmetries on both sides.
 
 \subsection{Conformal symmetry}

 The ${\cal N}=4$ SYM has an exact vanishing $\beta$-function and, therefore  is a CFT (invariant under the conformal group, which includes dilatations and special conformal transformations). On the gravity side, the space $AdS_5$ has the four-dimensional conformal group $SO(2,4)$  as isometry group. The $SO(2,4)$ group acts on the boundary  of $AdS_5$  as the conformal group in four dimensions. In particular, the dilatations act as:
 \beq
 (t,\vec x)\to \lambda  (t,\vec x)\,\,,
 \qquad
 z\to \lambda z\,\,.
 \eeq
 Therefore, the conformal symmetry is also realized on the gravity side if one identifies the field theory Minkowski spacetime as the boundary of $AdS_5$.

 It is also interesting to relate  the different scales on both theories. From the AdS metric we see (due to the $L^2/ z^2$ factor in front of the metric) that the proper distance $d$ on the bulk and the distance $d_{YM}$ on the Minkowski coordinates are related as:
\beq
d={L\over z} d_{YM}\,\,.
\eeq
For the energies this relation is inverted, since the energy is conjugated to the time. Then:
\beq
E={z\over L} E_{YM}\,\,.
\eeq
Thus, the high-energy limit (UV) in the field theory ($E_{YM}\to \infty$) corresponds to the near-boundary region $z\to 0$. Conversely, as $E_{YM}\to 0$ is equivalent to $z\to\infty$, the low energy limit (IR) corresponds to the near-horizon region $z\to\infty$. 
 
In a conformal theory, there exist excitations at arbitrary low energies. This corresponds to the fact that in AdS the geometry extends all the way to $z\to \infty$. On the contrary, in a non-conformal theory there should be a minimal scale and the geometry should end smoothly at some $z_0$. This is important when one generalizes the correspondence to non-conformal theories. There are two prominent examples of this generalized duality . The first case corresponds to the  confining theories with a mass gap $m$. In this case $z_0\sim 1/ m$. The second case are the theories at  finite temperature. Let $T$ denote the temperature. In this case the geometry is a black hole with a horizon and $z_0\sim 1/ T$.

 \subsection{ Supersymmetry}

Let us now see how one can match the supersymmetry on both sides of the correspondence. On the field theory side,  the  ${\cal N}=4$ SYM theory  is maximally supersymmetric. It has 32 fermionic supercharges generated by 4 sets of complex Majoranas $Q_{\alpha}^{A}$, $\bar Q_{\dot\alpha}^{A}$ ( $A=1\cdots, 4$). This corresponds to having ${\cal N}=4$ supersymmetry  in four dimensions. The $Q^{A}$ can be rotated with the group $SU(4)$,  which is the so-called $R$-symmetry group. Under this group the supercharges transform as:
\beq
Q_{\alpha}^{A}\to {\bf 4}\,\,,
\qquad
\bar Q_{\dot\alpha}^{A}\to \bar{\bf 4}\,\,.
\eeq
In addition, the theory has six scalars $\phi_1,\cdots \phi_6$, which transform as the fundamental representation ${\bf 6}$ of $SO(6)$. Notice the Lie algebra isomorphism $SU(4)\approx SO(6)$. 
 
The $AdS_5\times {\mathbb S}^5$ space is also a maximally supersymmetric solution of ten-dimensional supergravity. It has 32 Killing spinors, which correspond to the supercharges of ${\cal N}=4$ SYM. The rotational symmetry of the five-sphere is $SO(6)$, which can be identified  with the $R$-symmetry of ${\cal N}=4$ SYM. The directions along ${\mathbb S}^5$ correspond to the scalar fields on SYM. Thus, we have a perfect matching with the fields and symmetries of ${\cal N}=4$ SYM. Notice that the scalar fields of the field theory are extra coordinates on the gravity side and the isometries of the compact space are interpreted as internal rotations of the scalar fields and supercharges.

\subsection{Reduction on the ${\mathbb S}^5$}
\label{KK-reduction}

Let us now discuss further the role of the five-sphere in  the AdS dual of ${\cal N}=4$ SYM.  We begin by noticing that any field on $AdS_5\times {\mathbb S}^5$ can be reduced to a tower of fields on  $AdS_5$ by expanding  it in terms of the harmonics on ${\mathbb S}^5$:
\beq
\phi(x,\Omega)\,=\,\sum_{l}\,\phi_l (x)\,Y_l(\Omega)\,\,,
\label{KK_mode_expansion}
\eeq
with  $x$   being coordinates of  $AdS_5$, $\Omega$ are coordinates of  ${\mathbb S}^5$ and $Y_l(\Omega)$ are  spherical harmonics on ${\mathbb S}^5$. The gravity action, after reducing on  ${\mathbb S}^5$,  becomes:
\beq
S\,=\,{1\over 16\pi G_5}\,\int d^5 x\,\Big[\,{\cal L}_{grav}\,+\,{\cal L}_{matter}\,\Big]\,\,,
\eeq
where $G_5$ is the five-dimensional Newton constant, and the gravitational part of the ${\cal L}$ is:
\beq
{\cal L}_{grav}=\sqrt{-g}\,\Big[\,R\,+\,{12\over L^2}\,\Big]\,\,,
\eeq
which corresponds to a negative cosmological constant $\Lambda=-6/L^2$.
The five-dimensional Newton constant can be related to the 10d constant by considering the reduction of the Einstein-Hilbert term:
\beq
{1\over 16\pi G}\,\int d^5 x \,d^5\Omega \sqrt{-g_{10}}\,R_{10}\,\to\,
{L^5\Omega_5\over 16\pi G}\,\int d^5 x \, \sqrt{-g_{5}}\,R_{5}\,\,,
\eeq
where $\Omega_5=\pi^3$ is the volume of a unit  ${\mathbb S}^5$. Then it follows that  $G_5$ and $G$ can be related as:
\beq
G_5={G\over L^5\Omega_5}={G\over \pi^3 L^5}\,\,.
\eeq
Using the value of the ten-dimensional Newton constant $G$ written in (\ref{10d_Newton_constant}), we get that $G_5$ is given by:
\beq
G_5={\pi\over 2N^2}\,L^3\,\,.
\eeq
From this formula  and (\ref{holo_centralCharge}) we can compute the central charge of the SYM theory:
\beq
c_{SYM}={1\over 4}\,{L^3\over G_5}
\,=\,{N^2\over 2\pi}\,\,.
\eeq
Then, the large $N$ limit corresponds to having a large central charge, in agreement with our previous discussions on the validity domain of the gravitational description.

Let us now consider the reduction (\ref{KK_mode_expansion}) of  a massless scalar field in $AdS_5\times {\mathbb S}^5$. The Klein-Gordon equation in ten dimensions is:
\beq
\nabla^2 \phi\,=\,0\,\,.
\eeq
Since the metric factorizes into a  $AdS_5$ and ${\mathbb S}^5$ part, the D'Alembertian is additive
\beq
\nabla^2=\nabla^2_{AdS_5}+\nabla^2_{{\mathbb S}^5}\,\,,
\eeq
where $\nabla^2_{{\mathbb S}^5}$ is nothing but the quadratic Casimir operator in
$SO(6)$. The eigenvalues of $\nabla^2_{{\mathbb S}^5}$ acting on the spherical harmonics are:
\beq
\nabla^2_{{\mathbb S}^5}\,Y_l(\Omega)\,=\,-{C_l^{(5)}\over L^2}\,Y_l(\Omega)\,\,,
\eeq
where the  $C_l^{(5)}$ are given by:
\beq
C_l^{(5)}=l(l+4)\,\,,
\qquad
l=0,1,2,\cdots\,\,.
\eeq
Thus, the reduced  $AdS_5$ fields $\phi_l$ satisfy the massive Klein-Gordon equation
\beq
\nabla^2_{AdS_5}\,\phi_l\,=\,m_l^2\,\phi_l\,\,,
\qquad\qquad
m_l^2={l(l+4)\over L^2}\,\,.
\label{KK_masses}
\eeq
Therefore, we have a tower of massive fields $\phi_l$, with a particular set of masses, which originate, after  dimensional Kaluza-Klein (KK) reduction on the five-sphere,  from a single massless scalar field $\phi$ in ten-dimensions.  These fields should have a field theory dual in the ${\cal N}=4$ theory and the mass spectrum (\ref{KK_masses}) should have a counterpart on the field theory side. We will see that this is indeed the case in the next section.

\section{A scalar field in $AdS$}

We argued in section \ref{sec:1} that fields in AdS correspond to sources of operators on the field theory side and that we can learn about these dual operators by analyzing the dynamics of the sources in the curved space. In this section we study the simplest case of a scalar field in $AdS$. Accordingly,  let us consider the $AdS_{d+1}$ space in euclidean signature with metric:
 \beq
ds^2\,=\,{L^2\over z^2}\,[ dz^2+ \delta_{\mu\nu} \,
dx^{\mu}\,dx^{\nu}]\,\,.
\label{Euclidean_AdS_metric}
\eeq
The action of a scalar field $\phi$ in the $AdS_{d+1}$  space is:
\beq
S\,=\,-{1\over 2}\,\,\int d^{d+1} x \sqrt{g}\,\Big[
g^{MN}\,\partial_M\phi\,\partial_N\phi\,+\,m^2\,\phi^2\Big]\,\,.
\label{scalar_action_AdS}
\eeq
The equation of motion derived from the action (\ref{scalar_action_AdS}) is:
\beq
{1\over \sqrt{g}}\,\partial_M\,\Big(\,\sqrt{g}\,g^{MN}\partial_N\,\phi\Big)\,-\,m^2\,\phi\,=\,0\,\,.
\eeq 
More explicitly, after using the metric (\ref{Euclidean_AdS_metric}), this equation becomes :
 \beq
 z^{d+1}\,\partial_z\,\big(\,z^{1-d}\,\partial_z\phi\,\big)\,+\,z^2\,\delta^{\mu\nu}\,
 \partial_{\mu}\partial_{\nu}\,\phi\,-\,m^2\,L^2\,\phi\,=\,0\,\,.
 \label{fk_eq}
 \eeq
Let us perform the Fourier transform of $\phi$  in the $x^{\mu}$ coordinates:
\beq
\phi(z,x^{\mu})\,=\,\int {d^dk\over (2\pi)^d}\,\,
e^{ik\cdot x}\, f_k(z)\,\,.
\eeq
Then, the  equation of motion becomes:
\beq
z^{d+1}\,\partial_z\,\big(\,z^{1-d}\,\partial_z f_k\,\big)\,-\, k^2\,z^2\,f_k\,-\,
m^2\,L^2\,f_k\,=\,0\,\,.
\label{eom_phi}
\eeq
Let us solve (\ref{eom_phi}) near the boundary $z=0$. We put $f_k\sim z^{\beta}$ for some exponent $\beta$ and  keep the leading terms near $z=0$. Then, it is straightforward  to find that $\beta$ must satisfy the following quadratic expression:
\beq
\beta(\beta-d)\,-\,m^2\,L^2\,=\,0\,\,,
\eeq
whose solutions are:
\beq
\beta\,=\,{d\over 2}\pm\sqrt{{d^2\over 4}\,+\,m^2\,L^2}\,\,.
\eeq
Therefore, near $z\sim 0$ the function $f_k(z)$ behaves as :
\beq
f_k(z)\approx A(k)\,z^{d-\Delta}\,+\,B(k)\,z^{\Delta}\,\,,
\label{fk_near_z0}
\eeq
where $\Delta$  is given by:
\beq
\Delta={d\over 2}+\nu\,\,,
\qquad\qquad
\nu\,=\,\sqrt{{d^2\over 4}\,+\,m^2\,L^2}\,\,.
\label{Delta_nu}
\eeq
By performing the inverse Fourier transform we can write the expansion near the boundary in position space:
\beq
\phi(z,x)\approx A(x)\,z^{d-\Delta}\,+\,B(x)\,z^{\Delta}\,\,\,,
\qquad
z\to 0\,\,.
\label{scalar_near_boundary}
\eeq
Notice that $\Delta$ is real if $\nu\in {\mathbb R}$, which happens if the mass $m$ satisfies the  so-called Breitenlohner-Freedman (BF) bound:
\beq
m^2\ge -\Big({d\over 2L}\Big)^2\,\,.
\eeq
This means that $m^2$ can be negative (and the field $\phi$ can be tachyonic) but it must satisfy the BF bound. In what follows we will suppose that the BF bound is satisfied.
Moreover, notice that:
\beq
d-\Delta\le \Delta
\qquad
\Longleftrightarrow
\qquad
\nu=2\Delta-d\ge 0\,\,,
\eeq
which is obviously satisfied in the mass is above the BF bound. Then, the term behaving as $z^{d-\Delta}$ in (\ref{scalar_near_boundary}) is the dominant one as $z\to 0$. Let us take the boundary as $z=\epsilon$ and neglect the subdominant term. We have:
\beq
\phi(z=\epsilon, x)\approx \epsilon^{d-\Delta}\,A(x)\,\,. 
\label{phi_boundary}
\eeq
As $d-\Delta$ is negative if $m^2>0$, the leading term is typically divergent as we approach the boundary at $z=\epsilon\to 0$. In order to identify the QFT source $\varphi(x)$ from the boundary value of the field $\phi(z,x)$ we have to remove the divergences of the latter. We will simply do it by extracting the divergent multiplicative factor from (\ref{phi_boundary}), \ie\  the QFT source $\varphi(x)$  is identified with $A(x)$. Equivalently, we define:
\beq
\varphi(x)=\lim_{z\to 0}\,z^{\Delta-d}\,\phi(z,x)\,\,.
\eeq
Clearly, with this definition $\varphi(x)$ is always finite. Conversely we can write 
$\phi(z,x)=z^{d-\Delta}\,\varphi(x)$ at leading order.  In order to interpret the meaning of $\Delta$, let us look at the boundary action. If ${\cal O}$ is the operator dual to $\phi$, this action is given by:
\beq
S_{bdy}\sim \int d^dx\,\sqrt{\gamma_{\epsilon}}\,\,
\phi(\epsilon, x)\,{\cal O}(\epsilon, x)\,\,,
\eeq
where $\gamma_{\epsilon}=\Big({L\over \epsilon}\Big)^{2d}$ is the determinant of the induced metric at the $z=\epsilon$ boundary.  Then, by plugging $\phi(\epsilon,x)=\epsilon^{d-\Delta}\,\varphi(x)$ inside $S_{bdy}$, we get:
\beq
S_{bdy}\sim L^d\,\int d^dx\,\varphi(x)\,\epsilon^{-\Delta}\,
{\cal O}(\epsilon, x)\,\,.
\eeq
In order to make $S_{bdy}$ finite and independent of $\epsilon$ as $\epsilon\to 0$  we should require:
\beq
{\cal O}(\epsilon, x)=\epsilon^{\Delta}\,{\cal O}( x)\,\,.
\eeq
But, passing from $z=0$ to $z=\epsilon$ is a scale transformation in the QFT. Thus, $\Delta$ must be interpreted as the  mass scaling dimension of the dual operator ${\cal O}$. In other words, the behavior ${\cal O}(\epsilon, x)=\epsilon^{\Delta}\,{\cal O}( x)$ is the wave function renormalization of ${\cal O}$ as we move into the bulk. Similarly,  from the relation $\phi(\epsilon,x)=\epsilon^{d-\Delta}\,\varphi(x)$  it follows that  $d-\Delta$ is the mass scaling dimension of the source $\varphi$. Remember from (\ref{Delta_nu})  that:
\beq
\Delta\,=\,{d\over 2}+\sqrt{\Big({d\over 2}\Big)^2+m^2\,L^2}\,\,.
\label{dimension_mass}
\eeq
Then, if $m^2\ge -\Big({d\over 2L}\Big)^2$, the corresponding scaling dimension is real.
We consider three different situations, depending on the value of the mass $m$.  The first case we will study is  the one in which $m^2>0$.  In this case $\Delta>d$ and the corresponding operator ${\cal O}$ is called an irrelevant operator. By deforming the CFT with ${\cal O}$, we have the following term in the action:
\beq
\Delta S= \int d^dx\, (mass)^{d-\Delta}\,{\cal O}=
\int d^dx\, (mass)^{{\rm negative\,\,number}}\,{\cal O}\,\,. 
\eeq
When computing amplitudes with this action one gets that the effect of the interaction depends on $\Big({Energy\over mass}\Big)^{\alpha}$, for some $\alpha>0$. Then, the effects of this interaction can be neglected for low energies since they are suppressed by powers. Therefore, the interaction goes away in the IR but changes completely the UV of the theory.

When $m^2=0$ we have from (\ref{dimension_mass}) that  $\Delta=d$ and the corresponding  operator is called marginal. Finally, if  $m^2$ is negative and takes value if the range $-(d/ 2L)^2<m^2<0$, the dual operator it is called relevant, since it has $\Delta<d$ and changes the IR of the theory. 

\subsection{Application to the KK spectrum in $AdS_5\times {\mathbb S}^5$}

We have seen in section \ref{KK-reduction} that a massless scalar in ten-dimensions gives rise to the tower (\ref{KK_masses}) of massive KK scalars in $AdS_5$. We will now apply the formula (\ref{dimension_mass}) to these five-dimensional scalar fields. First of all, the dimension/mass relation for the case  $d=4$  is:
\beq
\Delta=2+\sqrt{4+(m\,L)^2}\,\,.
\label{Delta-mass-4d}
\eeq
Let us first consider a massless scalar, which corresponds to the s-wave ($l=0$) in (\ref{KK-reduction}). In this case (\ref{Delta-mass-4d}) with $m=0$ gives $\Delta=4$. Then, the QFT dual operator should be  a scalar operator of dimension 4. Since this s-wave operator is singlet under the $SO(6)$ symmetry of the ${\mathbb S}^5$, it must not contain the $\phi_i$  scalars of the dual ${\cal N}=4$ QFT. The only candidate with these characteristics is the glueball operator:
\beq
{\cal O}={\rm Tr}\big[ F_{\mu\nu}\,F^{\mu\nu}\big]\,\,.
\eeq
Notice that  dim$[\partial]$=dim$[A]=1$, so, indeed, dim($\,{\cal O}$)=4. For higher order KK  modes, the masses are $m^2 L^2=l(l+4)$ (see (\ref{KK_masses})). Then, the dimensions are:
\beq
\Delta_l=2+\sqrt{4+l(l+4)}\,=\,4+l\,\,. 
\eeq
In this case, the dual operator should transform under the corresponding representation of $SO(6)$ (a symmetric tensor with $l$ indices). One can construct such a tensor by multiplying ${\cal N}=4$ SYM scalar fields $\phi_i$ (they transform as vectors of $SO(6)$). Then, the natural operator dual to the $l^{{\rm th}}$ KK mode is:
\beq
{\cal O}_{i_1,\cdots, i_l}={\rm Tr}\big[\phi_{(i_1,\cdots, i_l)}
 F_{\mu\nu}\,F^{\mu\nu}\big]\,\,,
 \label{F_squared_adjoints}
 \eeq
with $\phi_{(i_1,\cdots, i_l)}$ being the traceless symmetric product of $l$ scalar fields $\phi_i$  of  ${\cal N}=4$ SYM.  As dim$[\phi]=1$, one can check immediately that the dimension of the operator 
${\cal O}_{i_1,\cdots, i_l}$ in  (\ref{F_squared_adjoints})  is indeed $4+l$, in agreement with the AdS/CFT result. It has been checked that this agreement can be extended to all the KK modes of 10d supergravity on  $AdS_5\times {\mathbb S}^5$ (including fermions, forms,...).

\subsection{Normalizable and non-normalizable modes}

The natural inner product for solutions of the Klein-Gordon equation in a curved space is:
\beq
(\phi_1,\phi_2)\,=\,-i\int_{\Sigma_t}\,
dz\,d^dx\,\sqrt{-g}\,g^{tt}\,
(\phi_1^*\,\partial_t\,\phi_2\,-\,\phi_2\,\partial_t\,\phi_1^*)\,\,,
\eeq
with $\Sigma_t$ being a constant-$t$ slice.  Let us consider in particular a field that behaves as $\phi\sim z^{\beta}$ near $z\approx 0$. In the $AdS_{d+1}$ metric $\sqrt{-g}\,g^{tt}\sim z^{-d+1}$. Then, $\phi\,\partial_t\,\phi\sim z^{2\beta}$ as $z\to 0$ and the integrand in the norm of $\phi$ behaves near $z=0$ as:
\beq
\sqrt{-g}\,g^{tt}\,
\phi\,\partial_t\,\phi\sim z^{2\beta-d+1}\,\,,
\eeq
which leads to a  convergent  integral if $2\beta-d+2>0$.

Let us now consider the two types of modes in (\ref{scalar_near_boundary}). We begin with the subleading modes $B(x)z^{\Delta}$.  In this case  the exponent  $\beta$ is $\beta= \Delta={d\over 2}+\nu$ and, therefore,  $2\beta-d+2=2\nu+2$, which is always positive. This mode is normalizable and can be considered as an element of the bulk Hilbert space and, given the equivalence with the dual theory, it should be identified with some state in the boundary theory. Next, we analyze the normalizability of the leading modes $A(x)z^{d-\Delta}$. 
For these modes $\beta=d-\Delta$ and, thus,   $2\beta-d+2=2(1-\nu)$, which is negative if $\nu\ge 1$. Notice that this condition is equivalent to $m^2 L^2\ge -{d^2\over 4}+1$. Then, the non-normalizable modes correspond to sources in the boundary theory. 
Interestingly,  for masses in the range $-{d^2\over 4}\le m^2\le  -{d^2\over 4}+1$ both types of modes are normalizable and they give rise to different Fock spaces of physical states. 

\subsection{Higher spin fields}

The results for the scalar field can be generalized to any $p$-form field, \ie\ to any antisymmetric tensor $A_{\mu_1\cdots \mu_p}$ with $p$ indices. For a $p$-form field of mass $m$, the dimension $\Delta$ of the dual operator is the largest root of the quadratic equation:
\beq
(\Delta-p)(\Delta+p-d)\,=\,m^2L^2\,\,.
\eeq
This equation can be solved to give the following dimension/mass relation:
\beq
\Delta={d\over 2}\,+\,\sqrt{
\Big({d-2p\over 2}\,\Big)^2\,+\,m^2 L^2}\,\,.
\label{dimension/mass-pform}
\eeq
In particular, for a massive vector field $A_{\mu}$, the previous formula (\ref{dimension/mass-pform}) with $p=1$ gives:
\beq
\Delta={d\over 2}\,+\,\sqrt{
\Big({d-2\over 2}\,\Big)^2\,+\,m^2 L^2}\,\,.
\label{dimension/mass-vector}
\eeq
Taking $m=0$ in (\ref{dimension/mass-vector}), we get $\Delta=d-1$, which is  the dimension of a conserved current $j_{\mu}$ in $d$ dimensions. Finally, one can prove that the dimension/mass relation for  a spin $1/2$ field is:
\beq
\Delta\,=\,{d\over 2}\,+\,|m\,L|\,\,.
\eeq

\section{Correlation functions}

Let us now see how one can compute correlation functions in Euclidean space from gravity. The objective is to obtain Euclidean correlation functions of the type:
 \beq
\langle {\cal O}(x_1)\cdots {\cal O}(x_n)\rangle\,\,.
\eeq
In field theory these correlators can be calculated from a generating function, which is 
obtained by perturbing the lagrangian by a source term:
\beq
{\cal L}\to {\cal L}+J(x)\,{\cal O}(x)\equiv {\cal L}+{\cal L}_J\,\,.
\eeq
The generating functional is just:
\beq
Z_{QFT}[J]=\Big\langle\,\, \exp[\int {\cal L}_J]\,\,\Big\rangle_{QFT}\,\,.
\eeq
The connected correlators are obtained from the functional derivatives of $Z$:
\beq
\Big\langle \prod_{i}{\cal O}(x_i)\Big\rangle=
 \prod_{i} {\delta\over \delta J(x_i)}\,\log Z_{QFT}[J]_{\big|J=0}
 \eeq
Let us consider now  any bulk field $\phi(z,x)$  fluctuating in $AdS$. Let $\phi_0(x)$ be the boundary value of $\phi$:
\beq
\phi_0(x)=\phi(z=0,x)=\phi |_{\partial AdS}(x)\,\,.
\eeq
The field $\phi_0$ is   related to a source for some dual operator ${\cal O}$ in the QFT. As we know the actual source is not the value of $\phi$ at $z=0$, which is typically divergent, but the limit:
\beq
\lim_{z\to 0}\,z^{\Delta-d}\phi(z,x)=\varphi(x)\,\,.
\label{renor-field}
\eeq
Then, the  AdS/CFT prescription for the generating functional is \cite{Gubser:1998bc,Witten:1998qj}:
\beq
Z_{QFT}[\phi_0]\,=\,
\Big\langle \exp\big[\int \phi_0\, {\cal O}\big]\,\Big\rangle_{QFT}\,=\,
Z_{gravity} [\phi\to \phi_0]\,\,,
\eeq
where $Z_{gravity} [\phi\to \phi_0]$ is the partition function (\ie\ the path integral) in the gravity theory evaluated over all functions  which have the value $\phi_0$ at the boundary of $AdS$:
\beq
Z_{gravity} [\phi\to \phi_0]=\sum_{\{\phi\to \phi_0\}} e^{S_{gravity}}\,\,.
\eeq
In the limit in which classical gravity dominates, one can substitute the sum by the term corresponding to the classical solution. In this case the generating function becomes:
\beq
Z_{QFT}[\phi_0]\,\approx\,
e^{S_{gravity}^{on-shell}[\phi\to\phi_0]}\,\,.
\eeq
One should be careful when evaluating the on-shell gravity action because it typically diverges and has to be renormalized following the procedure of holographic renormalization \cite{Henningson:1998gx,deHaro:2000xn} (see below and  the review \cite{Skenderis:2002wp}). Thus, the classical action must be substituted by a renormalized version, which will be denoted by $S_{grav}^{ren}$ and the generating functional becomes:
\beq
\log Z_{QFT}\,=\,S_{grav}^{ren}[\phi\to\phi_0]\,\,.
\eeq
Moreover, the $n$-point function can be obtained by computing the derivatives with respect to $\varphi=z^{\Delta-d}\phi$:
\beq
\langle {\cal O}(x_1)\cdots {\cal O}(x_n)\rangle=
{\delta^{(n)} S_{grav}^{ren}[\phi]\over \delta\varphi(x_1)\,\cdots \varphi(x_n)}\Bigg|_{\varphi=0}\,\,.
\eeq

\subsection{ One-point function}

It is also interesting to compute the one-point function of an operator ${\cal O}$  in the presence of the source $\varphi$:
\beq
\langle {\cal O}(x)\rangle_{\varphi}=
{\delta S_{grav}^{ren}[\phi]\over \delta\varphi(x)}\,\,.
\label{one-pt-source}
\eeq
Taking into account the relation between $\phi$ and $\varphi$ (eq. (\ref{renor-field})), we get:
\beq
\langle {\cal O}(x)\rangle_{\varphi}=\lim_{z\to 0}\,
z^{d-\Delta}\,\,{\delta S_{grav}^{ren}[\phi]\over \delta\phi(z,x)}\,\,.
\label{one-pt-fun-limit}
\eeq
The functional derivative of the classical on-shell action can be  computed in closed form. Indeed, let $S_{grav}$ be represented as:
\beq
S_{grav}=\int_{\cal M}\,\int dz\,d^dx {\cal L}[\phi,\partial\phi]\,\,,
\eeq
with ${\cal M}$ being a $(d+1)$-dimensional manifold whose boundary is located  at $z=0$. Under a general change $\phi\to\phi+\delta\phi$,  the classical action $ S_{grav}$ varies as: 
\beq
\delta S_{grav}= \int_{\cal M}\,\int dz\,d^dx
\Big[\,{\partial {\cal L}\over \partial \phi}\,\delta\phi\,+\,
{\partial {\cal L}\over \partial (\partial_{\mu} \phi)}\,\delta(\partial_{\mu}\phi)
\,\Big]\,\,.
\label{deltaS}
\eeq
Let us now  use in (\ref{deltaS}) that  $\delta[\partial_{\mu} \phi]=\partial_{\mu} (\delta\phi)$ and let us integrate by parts. We get:
\beq
\delta S_{grav}= \int_{\cal M}\,\int dz\,d^dx
\Big[\,\Big({\partial {\cal L}\over \partial \phi}-
\partial_{\mu} \Big({\partial {\cal L}\over \partial (\partial_{\mu} \phi)}\Big)\Big)
\delta\phi\,+\,\partial_{\mu} \Big({\partial {\cal L}\over \partial (\partial_{\mu} \phi)}
\delta\phi\Big)\,\Big]\,\,.
\eeq
The first term in the previous equation vanishes on-shell due to the Euler-Lagrange equations. As the boundary is at $z=\epsilon\to 0$, we can write:
\beq
\delta S_{grav}^{on-shell}= \int_{\epsilon}^{\infty} \int d^dx \,\,
\partial_{z} \Big({\partial {\cal L}\over \partial (\partial_{z} \phi)}
\delta\phi\Big)\,=\,-\int_{\partial M} d^dx\,
{\partial {\cal L}\over \partial (\partial_{z} \phi)}
\delta\phi\Big|_{z=\epsilon}\,\,.
\label{deltaS-onshell}
\eeq
Let us  next define $\Pi$ as:
\beq
\Pi=-{\partial {\cal L}\over \partial (\partial_{z} \phi)}\,\,,
\label{Pi}
\eeq
which is the canonical momentum if $z$ is taken as time.  Then (\ref{deltaS-onshell}) can be rewritten as:
\beq
\delta S_{grav}^{on-shell}=
\int_{\partial M} d^dx\,\Pi(\epsilon, x)\,\delta\phi(\epsilon, x)\,\,.
\eeq
Thus, it follows that:
\beq
{\delta  S_{grav}^{on-shell}\over \delta\phi(\epsilon, x)}\,=\,
\Pi(\epsilon, x)\,=\,-{\partial {\cal L}\over \partial (\partial_{z} \phi)}\,\,.
\eeq
In general the renormalized action can be written as:
\beq
S^{ren}\,=\, S_{grav}^{on-shell}+S_{ct}\,\,,
\label{S_ren}
\eeq
where $S_{ct}$ is the action of couterterms, defined at the boundary $z=\epsilon$. Let us define the renormalized momentum as:
\beq
\Pi^{ren}(z, x)={\delta S^{ren}\over \delta \phi(z,x)}\,\,.
\label{Pi_ren}
\eeq
Clearly, by taking $z=\epsilon$ in (\ref{Pi_ren}) and (\ref{S_ren}) we have:
\beq
\Pi^{ren}(\epsilon, x)=-{\partial {\cal L}\over \partial (\partial_{z} \phi(\epsilon,x))}+
{\delta S_{ct}\over \delta \phi(\epsilon,x)}\,\,.
\eeq
Therefore, from the definition of $\Pi^{ren}$ in (\ref{Pi_ren}) and eq. (\ref{one-pt-fun-limit}), we  obtain the one-point function of ${\cal O}$ in presence of the source $\varphi$  as the following limit in the AdS boundary:
\beq
\langle {\cal O}(x)\rangle_{\varphi}=\lim_{z\to 0}\,
z^{d-\Delta}\,\,\Pi^{ren}
(z,x)\,\,.
\label{one-point-Pi}
\eeq

\subsection{ Linear response theory}
\label{Linear_response}

The field theory path integral representation of the one-point function with  a source is:
\beq
\langle {\cal O}(x)\rangle_{\varphi}\,=\,
\int [D\psi]\,  {\cal O}(x)\, e^{S_E[\psi]+\int d^dy \,\varphi(y) {\cal O}(y) }\,\,,
\eeq
where $\psi$ denotes the fields of the QFT. Let us expand the exponent of this expression in a power series of the source $\varphi$ and let us keep the terms up to linear order:
\beq
\langle {\cal O}(x)\rangle_{\varphi}\,=\,\langle {\cal O}(x)\rangle_{\varphi=0}\,+\,
\int d^dy\,\langle {\cal O}(x)\, {\cal O}(y)\rangle\, \varphi(y)\,+\,\cdots\,\,.
\label{VEV_linear}
\eeq
Next, we define the euclidean two-point function $G_E(x-y)$ as:
\beq
G_E(x-y)\,=\,\langle {\cal O}(x)\, {\cal O}(y)\rangle\,\,.
\eeq
Then,  equation (\ref{VEV_linear}) can be rewritten as:
\beq
\langle {\cal O}(x)\rangle_{\varphi}\,=\,\langle {\cal O}(x)\rangle_{\varphi=0}\,+\,
\int d^dy\,G_E(x-y)\, \varphi(y)\,\,.
\eeq
We will consider normal-ordered observables such that 
$\langle {\cal O}(x)\rangle_{\varphi=0}$ vanishes. Notice that this always can be achieved by subtracting to ${\cal O}$  its  vacuum expectation value (VEV)  without source. Then, $\langle {\cal O}(x)\rangle_{\varphi}$ measures the fluctuations of the observable away from the expectation value, \ie\ the linear response of the system to the external perturbation and we can write:
\beq
\langle {\cal O}(x)\rangle_{\varphi}\,=\,\int d^dy\,G_E(x-y)\, \varphi(y)\,\,.
\eeq
In momentum space this expression can be written as:
\beq
\langle {\cal O}( k)\rangle_{\varphi}\,=\,
G_E( k)\,\varphi(k)\,\,,
\eeq
and, thus, we can obtain the two-point function in momentum space by dividing the one-point function by the source:
\beq
G_E(k)\,=\,{\langle {\cal O}( k)\rangle_{\varphi}
\over \varphi(k)}\,\,.
\label{G_E_varphi}
\eeq
In  the framework of the AdS/CFT correspondence,  since $\varphi=z^{\Delta-d} \phi(z,x)$ with $z\to 0$, we have the following formula for the two-point function in momentum  space:
\beq
G_E( k)\,=\,\lim_{z\to 0} z^{2(d-\Delta)}\,\,
{\Pi^{ren}(z, k)\over \phi(z, k)}\,\,.
\eeq

\subsection{Two-point function for a scalar field}

Let us apply the developments of the previous sections to the case in which the source $\phi$ is a scalar field in Euclidean $AdS_{d+1}$. We will assume that  the action of $\phi$ is:
\beq
S\,=\,-{\eta\over 2}\,\int dz\,d^d x\sqrt{g}\,\Big[
g^{MN}\partial_M\,\phi\,\partial_N\,\phi\,+\,m^2\,\phi^2\,\Big]\,\,,
\eeq
where $\eta$ is a normalization constant. In order to evaluate the on-shell action of $\phi$ we rewrite $S$ as follows:
\beq
S\,=\,-{\eta\over 2}\int dz\,d^d x \partial_{M} 
\Big[ \sqrt{g}\, \phi\, g^{MN}\partial_N\phi\Big]\,+\,
{\eta\over 2}\int dz\,d^d x\,\phi\,\sqrt{g}\,\Big[
{1\over \sqrt{g}}\,\partial_M\,\Big( \sqrt{g} g^{MN}\partial_N\phi\Big)\,-\,m^2\,\phi
\Big]\,\,.
\label{S_KGscalar}
\eeq
The second term  in (\ref{S_KGscalar}) vanishes when the equation of motion of $\phi$ is used. Then, the on-shell action is:
\beq
S^{on-shell}\,=\,-{\eta\over 2}\int dz\,d^d x \,\,\partial_{M} 
\Big[ \sqrt{g}\, \phi\, g^{MN}\partial_N\phi\Big]\,\,.
\label{S_KGscalar_onshell}
\eeq
Taking into account that the boundary is at $z=\epsilon$, which is the lower limit of the integration, the on-shell action (\ref{S_KGscalar_onshell}) can be written as:
\beq
S^{on-shell}\,=\,{\eta\over 2}\int d^d x \Big(\sqrt{g}\, \phi\, g^{zz}\partial_z\phi\Big)_{z=\epsilon}\,\,.
\eeq
Let us define the ``canonical momentum" $\Pi$ as in (\ref{Pi}):
\beq
\Pi\,\equiv\,-{\partial {\cal L}\over \partial (\partial_{z} \phi)}\,=\,
\eta\,\sqrt{g}\, g^{zz}\partial_z\phi\,\,.
\eeq
Then the on-shell action of $\phi$ becomes:
\beq
S^{on-shell}\,=\,{1\over 2}\,\int_{z=\epsilon}\,d^d x \,\,
\Pi(z,x)\,\phi(z,x)\,\,.
\eeq
Let us work in momentum space and  Fourier transform the field $\phi$ and the  canonical momentum $\Pi$:
\beq
\phi(z,x)\,=\,\int {d^dk\over (2\pi)^d}\,\,
e^{ik\cdot x}\, f_k(z)\,\,,
\qquad\qquad
\Pi(z,x)\,=\,\int {d^dk\over (2\pi)^d}\,\,
e^{ik\cdot x}\, \Pi_k(z)
\eeq
Then, Parseval identity allows to rewrite $S^{on-shell}$ as:
\beq
S^{on-shell}\,=\,{1\over 2}\int {d^dk\over (2\pi)^d}\,\,
\Pi_{-k}(z=\epsilon)\,f_k(z=\epsilon)\,\,.
\eeq
Recalling  that  the function $f_k(z)$ behaves near $z\sim 0$ as in (\ref{fk_near_z0}) and the definition of $\Pi$, we obtain:
\beq
\Pi(z,x)\approx\eta L^{d-1}\,
\Big[\,(d-\Delta)\,A(x)\,z^{-\Delta}\,+\,\Delta\, B(x)\, z^{\Delta-d}\,\Big]\,\,,
\qquad\qquad
(z\to 0)\,\,,
\eeq
which, in momentum space becomes:
\beq
\Pi_{-k}(z)\,\approx\,\eta L^{d-1}\,
\Big[\,(d-\Delta)\,A(-k)\,z^{-\Delta}\,+\,\Delta\, B(-k)\, z^{\Delta-d}\,\Big]\,\,,
\qquad\qquad
(z\to 0)\,\,,
\eeq
Let us use these results to compute the on-shell action, keeping the terms that do not vanish when $\epsilon\to 0$. We get:
\beq
S^{on-shell}\,=\,{\eta\over 2} L^{d-1}
\int {d^dk\over (2\pi)^d}\Big[
\epsilon^{-2\nu}\,(d-\Delta)\,A(-k)\,A(k)\,+\,d\,A(-k)\,B(k)\,\Big]\,\,.
\label{S_onshell_scalar_k}
\eeq
Notice that the first term in (\ref{S_onshell_scalar_k}) is divergent. We will now find a counterterm to renormalize this divergence of the on-shell action. It  must be a local quadratic functional defined at the boundary of $AdS$. The natural candidate would be a term proportional to:
\beq
\int_{\partial AdS} d^d x\, \sqrt{\gamma}\,\,\phi^2 (\epsilon, x)\,\,,
\eeq
where $\gamma$ is the determinant of the induced metric  $\gamma_{\mu\nu}$  on the boundary:
\beq
ds^2_{z=\epsilon}=\gamma_{\mu\nu}\,dx^{\mu}\,dx^{\nu}\,=\,
{L^2\over \epsilon^2}\,\,\delta_{\mu\nu} dx^{\mu}\,dx^{\nu}\,\,.
\eeq
It is easy to prove  that:
\beq
\int_{\partial AdS} d^d x\, \sqrt{\gamma}\,\phi^2 (\epsilon, x)\,=\,
L^d\,\int {d^dk\over (2\pi)^d}\Big[
\epsilon^{-2\nu}\,A(-k)\,A(k)+2\,A(-k)\,B(k)\,\Big]\,\,,
\eeq
Let us adjust the coefficient of the counterterm action in such a way that the leading divergence is cancelled. It is immediate to check that this counterterm action must be:
\beq
S_{ct}\,=\,-{\eta\over 2}\,\, {d-\Delta\over L}\,\,
\int_{\partial AdS} d^d x\, \sqrt{\gamma}\,\phi^2\,\,,
\eeq
or, equivalently, in momentum space:
\beq
S_{ct}\,=\,-{\eta\over 2}\,\, (d-\Delta)\, L^{d-1}\,\,
\int {d^dk\over (2\pi)^d}\Big[
\epsilon^{-2\nu}\,A(-k)\,A(k)\,+\,2\,A(-k)\,B(k)\,\Big]\,\,.
\label{Sct_scalar_k}
\eeq
The renormalized action $S^{ren}=S^{on-shell}+S_{ct}$  can be obtained by adding (\ref{S_onshell_scalar_k}) and (\ref{Sct_scalar_k}), with the result:
\beq
S^{ren}\,=\,{\eta\over 2}\, L^{d-1}\,(2\Delta-d)\,
\int {d^dk\over (2\pi)^d}\,A(-k)\,B(k)\,\,.
\label{Sren}
\eeq
In order to extract the one-point function from $S^{ren}$ we have to compute the functional derivative with respect to $\varphi(x)$(recall that $\varphi(x)=A(x)$). However, the coefficient $B(x)$ also depends functionally on $A(x)$. To illustrate this point,  let us represent $f_k(z)$ for arbitrary $z$ (not necessarily small) as:
\beq
f_k(z)\,=\,A(k)\,\phi_1(z,k)\,+\,B(k)\,\phi_2(z,k)\,\,,
\label{general_fk}
\eeq
where $\phi_1(z,k)$ and $\phi_2(z,k)$ are  independent solutions of the equation satisfied by $f_k(z)$, normalized in such a way that for $z\to 0$ they behave as:
\beq
\phi_1(z,k)\approx z^{d-\Delta}
\,\,,
\qquad\qquad
\phi_2(z,k)\approx z^{\Delta}\,\,.
\eeq
(the explicit expressions of  $\phi_1(z,k)$ and $\phi_2(z,k)$ are given below). To determine completely $\phi$ we have to impose regularity conditions in the deep IR $z\to\infty$. As we will see soon, this fixes uniquely the ratio $B/A$ to a value which is independent of the value of the field at the boundary $z=0$. Let us denote  this ratio by $\chi$:
\beq
\chi\,=\,{B\over A}\,\,.
\label{chi_def}
\eeq
Clearly, we can write the renormalized action (\ref{Sren}) as:
\beq
S^{ren}\,=\,{\eta\over 2}\, L^{d-1}\,(2\Delta-d)\,
\int {d^dk\over (2\pi)^d}\,\chi(k) 
\varphi(k)\,\varphi(-k)\,\,,
\eeq
where we have used the fact that $\phi(k)=A(k)$. Then, as the functional derivative 
${\delta\over \delta \varphi(x)}$ is equivalent to $(2\pi)^d {\delta\over \delta \varphi(-k)}$ in momentum space, we have:
\beq
\langle {\cal O}( k)\rangle_{\varphi}\,=\,(2\pi)^d \,{\delta S^{ren}\over \delta \varphi(-k)}\,=\,
\eta\,L^{d-1}\,(2\Delta-d)\,\chi(k) 
\varphi(k)\,\,.
\label{one-point_chi}
\eeq
Taking into account the definition of $\chi$ in (\ref{chi_def}) and that $2\Delta-d=2\nu$, we can write (\ref{one-point_chi}) as:
\beq
\langle {\cal O}( k)\rangle_{\varphi}\,=\,2\nu\,\eta \,L^{d-1}\, B(k)\,\,.
\eeq
Thus, the subleading contribution near the boundary ($B(k)$) determines the VEV of the operator. The two-point function $G_E(k)$ can be obtained by dividing by the source (see (\ref{G_E_varphi})):
\beq
G_E(k)\,=\,2\nu\,\eta \,L^{d-1}\,{B(k) \over A(k)}\,\,.
\eeq
Let us now calculate explicitly $A(k)$ and $B(k)$. We first define the function $g_k(z)$ as:
\beq
f_k(z)\,=\,z^{{d\over 2}}\,g_k(z)\,\,.
\eeq
Then, one can easily check from (\ref{fk_eq}) that $g_k(z)$ satisfies the equation:
\beq
z^2\,\partial_z^2\,g_k\,+\,z\,\partial_z\,g_k\,-\,(\nu^2\,+\,k^2\,z^2)g_k\,=\,0\,\,.
\label{g_k_equation}
\eeq
Eq. (\ref{g_k_equation}) is just  the modified Bessel equation, whose two independent solutions can be taken to be $g_k=I_{\pm \nu}(kz)$, where $I_{\pm \nu}$ are modified Bessel functions. Thus, the two independent solutions for $f_k(z)$ are:
\beq
z^{{d\over 2}}\,I_{\pm \nu}(kz)\,\,.
\eeq
Notice that for $z\to 0$ the modified Bessel functions behave as:
\beq
I_{\pm \nu}(z)\approx {1\over \Gamma(1\pm \nu)}\, \Big({z\over 2}\Big)^{\pm\nu}
\,\,,
\qquad (z\to 0)\,\,.
\eeq
Then, we can take $\phi_1(z,k)$ and $\phi_2(z,k)$  in (\ref{general_fk}) as:
\beq
\phi_1(z,k)\,=\,\Gamma(1- \nu)\,\Big({k\over 2}\Big)^{\nu}\,z^{{d\over 2}}\,I_{-\nu}(kz)
\,\,,
\qquad\qquad
\phi_2(z,k)\,=\,\Gamma(1+ \nu)\,\Big({k\over 2}\Big)^{-\nu}\,z^{{d\over 2}}\,I_{\nu}(kz)\,\,.
\eeq
One can check that these functions have, indeed, the correct behavior when $z\to 0$. Then, by using (\ref{general_fk}) we get:
\beq
f_k(z)\,=\,z^{{d\over 2}}\,\Big[
\Gamma(1- \nu)\,\Big({k\over 2}\Big)^{\nu}\,A(k)\,I_{-\nu}(kz)\,+\,
\Gamma(1+ \nu)\,\Big({k\over 2}\Big)^{-\nu}\,B(k)\,I_{\nu}(kz)\Big]\,\,.
\eeq
Let us now impose that $f_k(z)$ is finite when $z\to \infty$. This condition determines a precise relation between the coefficients $A(k)$ and $B(k)$, as already mentioned above. When $z\to \infty$, the functions $I_{\pm \nu}(z)$ behave as:
\beq
I_{\pm \nu}(z)\approx {e^{z}\over \sqrt{2\pi z}}
\,\,\,,
\qquad\qquad 
(z\to\infty)\,\,.
\eeq
Then, after changing $z\to kz$, we get for large $z$:
\beq
f_k(z)\approx {z^{{d\over 2}}\,e^{kz}\over \sqrt{2\pi kz}}
\Big[
\Gamma(1- \nu)\,\Big({k\over 2}\Big)^{\nu}\,A(k)\,+\,
\Gamma(1+ \nu)\,\Big({k\over 2}\Big)^{-\nu}\,B(k)\Big]\,\,,
\eeq
which diverges when $z\to\infty$ unless the coefficient in brackets vanishes. Then, we must require:
\beq
{B(k)\over A(k)}\,=\,-{\Gamma(1-\nu)\over \Gamma(1+\nu)}\,
\Big({k\over 2}\Big)^{2\nu}\,=\,{\Gamma(-\nu)\over \Gamma(\nu)}\,
\Big({k\over 2}\Big)^{2\nu}\,\,.
\eeq
Using this result we can compute the  euclidean two-point function. We get:
\beq
G_E(k)\,=\,2\nu\,\eta \,L^{d-1}\,
{\Gamma(-\nu)\over \Gamma(\nu)}\,
\Big({k\over 2}\Big)^{2\nu}\,\,.
\eeq
Let us write this result in position space. The relation between $G_E(x)$ and $G_E(k)$ is: 
\beq
G_E(x)=\int {d^dk\over (2\pi)^d}\,e^{ikx}\,G_E(k)\,\,.
\eeq
We now use the formula:
\beq
\int {d^dk\over (2\pi)^d}\,e^{ikx}\,k^n\,=\,
{2^n\over \pi^{{d\over 2}}}\,\,
{\Gamma\big({d+n\over 2}\big)\over 
\Gamma\big(-{n\over 2}\big)}\,
{1\over |x|^{d+n}}\,\,,
\eeq
to obtain the correlator in momentum space:
\beq
\langle {\cal O}(x) {\cal O}(0)\rangle=
{2\nu \eta  L^{d-1}\over \pi^{{d\over 2}}}\,
{\Gamma\big({d\over 2}+\nu\big)\over 
\Gamma\big(-\nu\big)}\,
{1\over |x|^{2\Delta}}\,\,.
\label{2-point-position}
\eeq
The behavior $\langle {\cal O}(x) {\cal O}(0)\rangle\sim |x|^{-2\Delta}$ in (\ref{2-point-position}) confirms that $\Delta$ is, indeed,  the scaling dimension of the operator ${\cal O}(x)$.

\section{Quark-antiquark potential}

Let us consider an external charge moving along a closed curve ${\cal C}$ in spacetime in QED. The action for such a charge is:
\beq
S_{{\cal C}}\,=\,\oint_{{\cal C}}\,A_{\mu}\,dx^{\mu}\,\,.
\eeq
Adding this term to the action is equivalent to insert in the path integral the quantity:
\beq
e^{iS_{{\cal C}}}\,=\,e^{i\oint_{{\cal C}}\,A_{\mu}\,dx^{\mu}}\,
\equiv \,W({\cal C})\,\,,
\eeq
where $W({\cal C})$ is the so-called Wilson loop, which is just the holonomy of the gauge field $A_{\mu}$ along the closed curve ${\cal C}$. Notice that $W({\cal C})$ is the Aharonov-Bohm phase factor due to the propagation of a quark along the closed curve ${\cal C}$. The non-abelian analogue of the previous formula is:
\beq
W({\cal C})={\rm Tr}\,P\,\exp\Big[i\oint_{{\cal C}}\,A_{\mu}\,dx^{\mu}\Big]\,\,,
\eeq
where $A_{\mu}=A_{\mu}^a\,T^a$, ${\rm Tr}$ is the trace over the group indices and $P$ is the path ordering operator. Notice that $\langle W({\cal C}) \rangle$ can be regarded as the amplitude for a creation of a $q\bar q$ pair which propagates and is annihilated afterwards. 

\begin{figure}[ht]
\center
\includegraphics[width=0.6\textwidth]{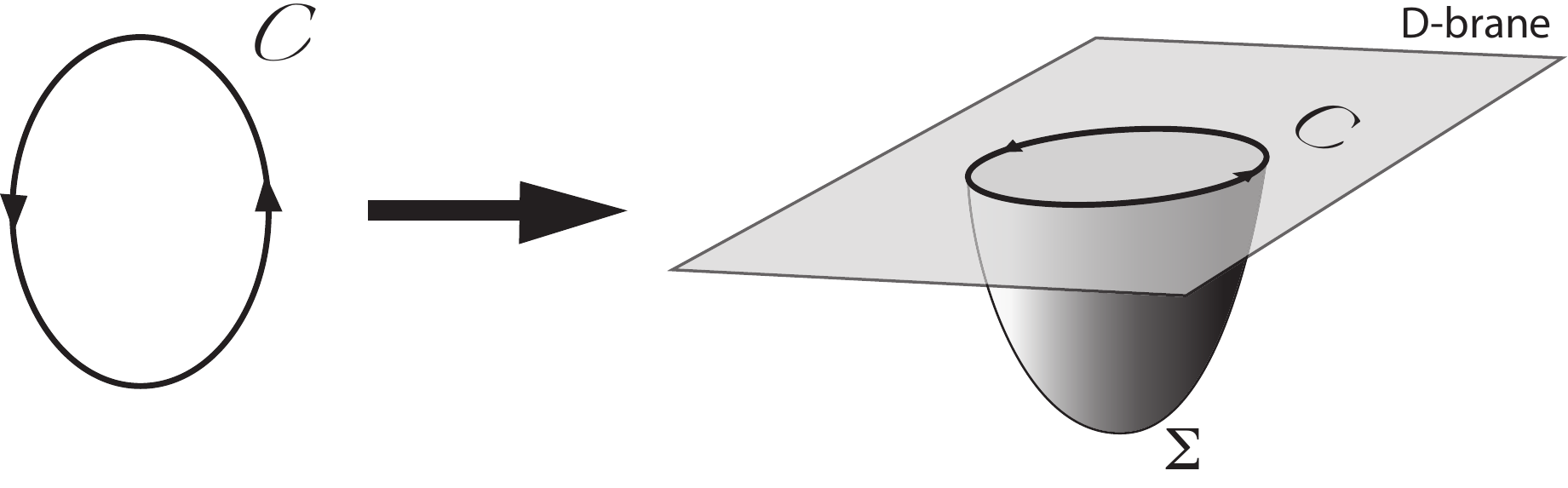}
\qquad
\caption{A Wilson loop along a closed curve ${\cal C}$ is holographically realized as the boundary of a surface $\Sigma$.  } 
\label{holoWilson}
\end{figure}

In particular, we can take a rectangular Wilson loop in Euclidean space.  Let $T$ and $d$ be the sizes of the rectangle. In this case the Wilson loop is the amplitude for the propagation of a $q\bar q$ pair separated a distance $d$, which can be calculated by means of the hamiltonian formalism. Actually, when $T\to\infty$ one has:
\beq
\lim_{T\to\infty}\,\langle W({\cal C})\rangle\sim e^{-T\,E(d)}\,\,,
\eeq
where $E(d)$ is the energy of a $q\bar q$ pair separated a distance $d$. 
In a confining theory the energy grows linearly with the $q\bar q$ distance $d$:
\beq
E(d)\approx \sigma\, d\,\,,
\qquad
\sigma\to {\rm constant}\,\,.
\eeq
Then, the Wilson loop in a confining theory satisfies the so-called area law:
\beq
\lim_{T\to\infty}\,\langle W({\cal C})\rangle\sim e^{-\sigma\,T\,d}\sim 
e^{-\sigma\,({\rm area \,enclosed\, by \,the\, loop})}\,\,.
\eeq
Let us see how we can compute  VEVs of Wilson loops in the AdS/CFT correspondence \cite{Maldacena:1998im,Rey:1998ik}. Recall that the endpoint of an open string ending on a D-brane is dual to a quark. Thus, in string theory, the Wilson loop is represented by an open string whose worldsheet  $\Sigma$ has a boundary $\partial\Sigma$ which is in a D-brane and describes the curve ${\cal C}$, as shown in figure \ref{holoWilson}. Then:
\beq
\langle W({\cal C})\rangle=Z_{string} \big(\partial\Sigma={\cal C}\big)\,\,.
\eeq
To have infinitely heavy (non-dynamical) quarks we push the D-brane to the AdS boundary and, therefore, $\partial\Sigma={\cal C}$ lies within the AdS boundary $z=0$. Then,  in the 't Hooft limit, we have:
\beq
Z_{string} \big(\partial\Sigma={\cal C}\big)\,=\,e^{-S({\cal C})}\,\,,
\eeq
where $S({\cal C})$ is the on-shell extremal Nambu-Goto action for the string worldsheet satisfying the boundary condition that $\Sigma$  ends on the curve ${\cal C}$. The Wilson loop becomes simply:
\beq
\langle W({\cal C})\rangle=e^{-S({\cal C})}\,\,.
\eeq

We will now apply these ideas to the holographic calculation of the rectangular Wilson loop and the corresponding $q\bar q$ potential energy in ${\cal N}=4$ SYM. With this purpose, 
let us consider a string ending on the boundary of $AdS_5$, where it extends along one of the gauge theory coordinates (say $x^1\equiv x$). 
We shall parameterize the embedding of the string in $AdS_5$ by a function $z=z(x)$ and we will denote by $z'$ the derivative of $z$ with respect to $x$. 
The induced metric in Euclidean signature on the string worldsheet is:
\beq
ds^2\,=\, {L^2\over z^2}\,\big[\,dt^2+(1+z'^{\,2})\,dx^2\,\big]\,\,.
\eeq
Therefore, the Nambu-Goto action for the string is:
\beq
S\,=\,{1\over 2\pi\alpha'}\,\int dt\,\int dx\,\sqrt{g}\,=\,
{T L^2\over 2\pi\alpha'}\,\int dx 
{\sqrt{1+z'^{\,2}}\over z^2}\,\,,
\label{action_hanging_string}
\eeq
where $T=\int dt$. The action (\ref{action_hanging_string}) corresponds to a problem in classical mechanics where $x$ is the ``time"  and the  lagrangian is given by:
\beq
{\cal L}\,=\,{T L^2\over 2\pi\alpha'} \,\,
{\sqrt{1+z'^{\,2}}\over z^2}\,\,.
\eeq
Since this lagrangian does not depend on the  ``time" $x$ , the ``energy" is conserved, which means that: 
\beq
z'\,{\partial {\cal L}\over \partial z'}\,-\,{\cal L}\,=\,{\rm constant}\,\,.
\label{energy_conservation}
\eeq
By computing the derivative of ${\cal L}$ appearing on the left-hand side of (\ref{energy_conservation}), the conservation law written above can be proved to be equivalent to the following first integral of the equation of motion:
\beq
 z^2\,\sqrt{1+z'^{\,2}}\,=\,{\rm constant}\,\,.
 \label{first_integral}
 \eeq
For the hanging string configuration we are interested in, the boundary conditions that the function $z(x)$ must satisfy are:
\beq
z\big(x=-d/2\big)=z\big(x=d/2\big)=0\,\,,
\label{bc_hanging_string}
\eeq
where $d$ is the quark-antiquark separation. 
Clearly, there is a value of $x$ for which $z$ is maximal. By symmetry this value is just 
$x=0$ and, since $z(x)$ has a maximum implies that  $z'(x=0)=0$. Let $z_*$ be the maximal value of $z$, \ie\ $z_*=z(x=0)$. Then, the constant appearing in the conservation law is just $z_*^2$ and we have:
\beq
z'=\pm {\sqrt{z_*^4-z^4}\over z^2}\,\,,
\label{zprime_hanging_string}
\eeq
where the two signs correspond to the two sides of the hanging string. The equation written above can be integrated immediately to give $x$ as a function of $z$:
\beq
x\,=\,\pm \int_{z_*}^{z}\,{\xi^2\over \sqrt{z_*^4-\xi^4}}\,d\xi\,\,.
\eeq
By introducing a new rescaled variable $y$ by means of the relation $\xi=z_* y$, one can write:
\beq
x\,=\,\pm z_*\,\int_{1}^{{z\over z_*}}\,
{y^2\over \sqrt{1-y^4}}\,dy\,\,.
\eeq
By imposing the boundary conditions (\ref{bc_hanging_string}), we can obtain the quark-antiquark separation as a function of $z_*$:
\beq
{d\over 2}\,=\,z_*\,\int_{0}^{1}\,{y^2\over \sqrt{1-y^4}}\,dy\,\,.
\label{d_integral}
\eeq
The integral over $y$ in (\ref{d_integral}) can be computed exactly and one gets:
\beq
z_*\,=\,{d\over  2\sqrt{2}\,\pi^{{3\over 2}}}\,\,
\Big(\Gamma\Big({1\over 4}\Big)\Big)^2\,\,.
\eeq
Then, it follows that $d$ and $z_*$ are proportional.  Let us now compute the on-shell action for the string, from which we will obtain the quark-antiquark potential. By plugging the conservation law (\ref{first_integral}) inside the expression  of the action in (\ref{action_hanging_string}), we get:
\beq
S\,=\,{T\,L^2 z_*^2\over 2\pi\alpha'}\,\int {dx\over z^4}\,\,.
\label{on-shell_action_hanging_string}
\eeq
Let us change variables in the integral  (\ref{on-shell_action_hanging_string}) from $x$ to $z$. The jacobian for this change of variables  is:
\beq
{dx\over dz}\,=\,{1\over z'}\,=\,{z^2\over \sqrt{z_*^4-z^4}}\,\,,
\eeq
where we have taken into account the value of $z'$ written in (\ref{zprime_hanging_string}). Taking into account that the variable $z$ is double-valued, the total action is:
\beq
S\,=\,2\times {T L^2 z_*^2\over 2\pi\alpha'} \int_{\epsilon}^{z_*}
{dz\over z^2\sqrt{z_*^4-z^4}}\,\,.
\eeq
Let us next work on a rescaled variable $y$, related to $z$ as $z=z_* y$. Then, the on-shell action becomes:
\beq
S\,=\,{T L^2 \over \pi\alpha'z_*}\,I_{\epsilon}\,\,,
\eeq
where $I_{\epsilon}$ is the following integral:
\beq
I_{\epsilon}\,=\,\int_{\epsilon/z_*}^{1}\,
{dy\over y^2\sqrt{1-y^4}}\,\,.
\label{I_epsilon}
\eeq
The integral $I_{\epsilon}$ in (\ref{I_epsilon})  diverges when $\epsilon\to 0$. Indeed, one can prove that, for small $\epsilon$:
\beq
I_{\epsilon}\,=\,-{\pi^{{3\over 2}}\,\sqrt{2}\over \Big(\Gamma\Big({1\over 4}\Big)\Big)^2}\,+\,{z_*\over \epsilon}\,\,.
\eeq
The  on-shell action $S$ is just $T\,E$, where $E$ is the energy of the quark-antiquark. It follows that $E$ is given by:
\beq
E\,=\,-{4\pi^2 L^2\over \Big(\Gamma\Big({1\over 4}\Big)\Big)^4\,\alpha'}\,{1\over d}\,+\,
{L^2\over \pi\alpha'}\,\,{1\over \epsilon}\,\,.
\label{unregularized_energy_string}
\eeq
The divergent term  in (\ref{unregularized_energy_string}) corresponds to the quark  and antiquark masses (which are considered to be infinitely large in the static limit). To check this, let us compute the euclidean action of a single string which goes straight  from the boundary $z=\epsilon$ to $z=\infty$ at fixed $x$. This configuration can be described by using $t$ and $z$ as worldvolume coordinates. The induced metric is obtained by keeping constant the $x$ coordinate and is given by:
\beq
ds^2\,=\,{L^2\over z^2}\,(dt^2+dz^2)\,\,.
\eeq
The on-shell Nambu-Goto action for this configuration is:
\beq
S_{|}\,=\,{T\over 2\pi\alpha'}\,\,\int_{\epsilon}^{\infty} {L^2\over z^2} dz^2\,=\,
{TL^2\over  2\pi\alpha'\,\epsilon}\,\,.
\eeq
Therefore, the energy for a configuration of two straight parallel strings in $AdS_5$ is:
\beq
E_{|\,|}\,=\,2\times {L^2\over  2\pi\alpha'\,\epsilon}\,\,,
\label{E_parallel}
\eeq
which is equal to the divergent term in (\ref{unregularized_energy_string}), as claimed. 
The quark-antiquark potential $V_{q\bar q}$ is obtained by subtracting the divergent contribution due to the quark masses:
\beq
V_{q\bar q}\,=\,E\,-\,E_{|\,|}\,\,.
\eeq
By subtracting (\ref{unregularized_energy_string}) and (\ref{E_parallel}), one gets:
\beq
V_{q\bar q}\,=\,-{4\pi^2 L^2\over \Big(\Gamma\Big({1\over 4}\Big)\Big)^4
\alpha'}\,\,{1\over d}\,\,.
\label{V_q-barq_gravity}
\eeq
Let us write this result in terms of gauge theory quantities. Recall that 
$L^2=\sqrt{N^2 g_{YM}}\,\alpha'$ or, equivalently $L^2=\sqrt{\lambda}\,\alpha'$. Then, (\ref{V_q-barq_gravity}) can be rewritten as:
\beq
V_{q\bar q}\,=\,-{4\pi^2 \sqrt{\lambda}\over \Big(\Gamma\Big({1\over 4}\Big)\Big)^4 }\,\,{1\over d}\,\,.
\label{q_barq_gauge}
\eeq
The Coulombic $1/d$ dependence in (\ref{q_barq_gauge}) is consequence of the conformal invariance of the theory. The non-analytic dependence on the coupling $\lambda$ is a non-perturbative effect. In field theory this non-analyticity  results from the summation of an infinite number of Feynman diagrams. It is interesting to compare (\ref{q_barq_gauge}) with the perturbative $q\bar q$ potential, valid for small $\lambda$, which is given by:
\beq
V_{q\bar q}\,=\,-{\pi\lambda\over d}\,\,.
\eeq
Remarkably we have been able to find a non-perturbative result in a interacting quantum field theory by studying the catenary curve for a hanging string in classical gravity. 

In the next two sections we will generalize the result (\ref{q_barq_gauge}) to the case of ${\cal N}=4$ SYM at finite temperature and to the gravity dual of a theory in which the quarks are confined. In the former case we have to study a string hanging from the boundary of an anti-de-Sitter black hole, whereas in the latter we must deal with a geometry obtained from the AdS black hole by a double analytic continuation.

\subsection{Quark-antiquark potential at finite temperature}

Let us suppose that we put our gauge theory in a thermal bath at a temperature $T$. In the holographic correspondence the heat bath is dual to a bulk geometry with an event horizon (a black hole) and the Hawking temperature of the black hole is identified as the temperature of heat bath.  In particular, the dual of ${\cal N}=4$ SYM at finite temperature is a black hole in $AdS_5$, whose metric  in euclidean space is (see below):
\beq
ds^2\,=\,{L^2\over z^2}\,\Big[\,f(z)\,dt^2\,+\,d\vec x^{\,2}\,+\,{dz^2\over f(z)}\,\Big]\,\,,
\eeq
 with $f(z)$ being the following function:
 \beq
 f(z)\,=\,1\,-\,{z^4\over z_0^4}\,\,.
 \eeq
The parameter $z_0$ corresponds to the position of the horizon and is related to the black hole temperature $T$ as $T=(\pi z_0)^{-1}$ (see section \ref{section_AdS_BH} below).  In the context of the holographic duality one naturally identifies the black hole temperature $T$ with the temperature of the gauge theory.

\begin{figure}[ht]
\center
\includegraphics[width=0.30\textwidth]{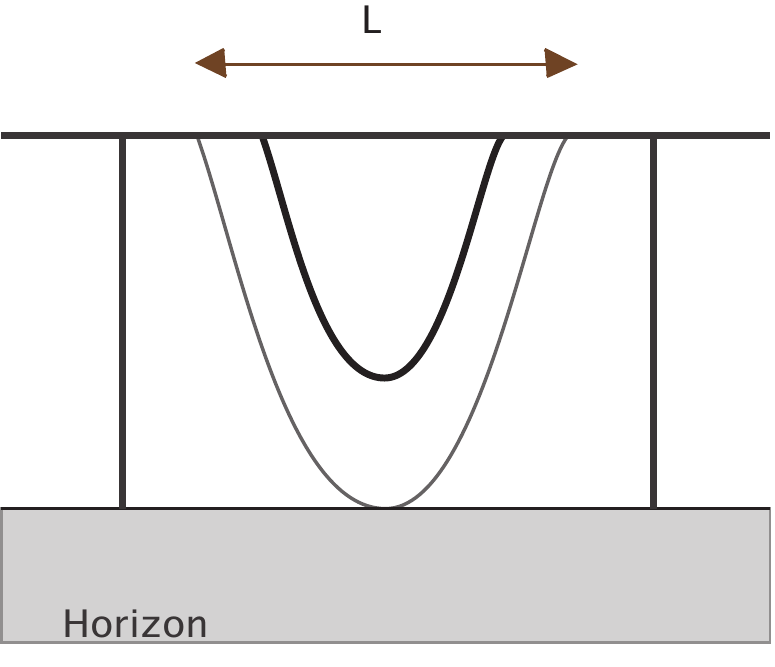}
\qquad
\includegraphics[width=0.60\textwidth]{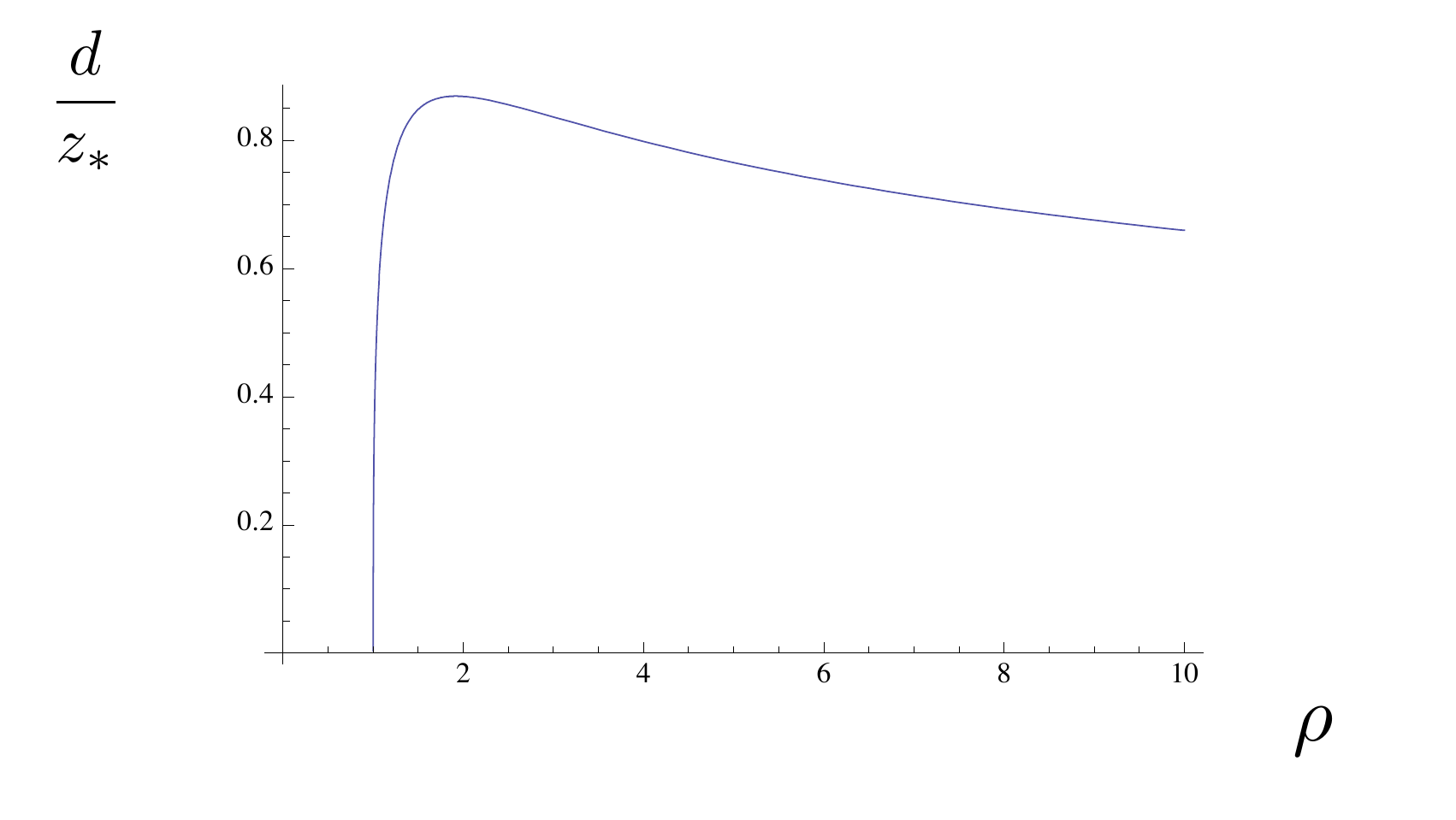}
\caption{In a black hole metric, the string ends on the horizon when the $q\bar q$ separation is large enough. Correspondingly, the distance $d$ reaches a maximum value, as the plot on the right shows.  } 
\label{Wilson_finiteT}
\end{figure}

As in the zero temperature case, we will parametrize the embedding of the string by a function $z=z(x)$. The induced metric is:
\beq
ds^2\,=\, {L^2\over z^2}\,\Big[\,f\,dt^2+\Big(1+{z'^{\,2}\over f}\Big)\,dx^2\,\Big]\,\,.
\eeq
The Nambu-Goto action for this embedding becomes:
\beq
S\,=\,{\tau L^2\over 2\pi\alpha'}\,\int dx 
{\sqrt{f(z)+z'^{\,2}}\over z^2}\,\,,
\eeq
where $\tau=\int dt$. Following the same steps as in the zero temperature case, it is straightforward to derive the following first integral of the equation of motion:
\beq
{z^2\,\sqrt{f(z)+z'^{\,2}}\over f(z)}\,=\,{\rm constant}\,=\,{z_*^2\over \sqrt{f(z_*)}}\,\,.
\label{first_integral_BH}
\eeq
From (\ref{first_integral_BH}) we obtain readily the value of $z'$:
\beq
z'\,=\,\pm \sqrt{{f(z)\over f(z_*)}}\,\,
{\sqrt{z_*^4-z^4}\over z^2}\,\,.
\eeq
Let us now define the constant $\rho$ as follows:
\beq
\rho\,\equiv \Big({z_0\over z_*}\Big)^4\,\,.
\label{rho_def}
\eeq
Then, we have:
\beq
x\,=\,\pm z_*\,\sqrt{\rho-1}\,
\int_{1}^{{z\over z_*}}\,
{y^2 dy\over \sqrt{(1-y^4)(\rho-y^4)}}\,\,,
\eeq
and the quark-antiquark distance $d$ is equal to:
\beq
d\,=\,2 z_*\,\sqrt{\rho-1}\,
\int_{0}^{1}\,
{y^2 dy\over \sqrt{(1-y^4)(\rho-y^4)}}\,\,.
\eeq
Notice that these expressions become the ones at zero temperature as $\rho\to\infty$, as it should. Moreover, as  $z_*\to z_0$, $\rho\to 1$ and  the distance $d\to 0$ (see figure \ref{Wilson_finiteT}). Actually, by varying $\rho$ one can see that there is a maximal value of $d$ ($d_{max}\sim z_0$) and the $q\bar q$ bound state becomes unbound due to thermal screening, in agreement with the behavior expected from the point of view of the gauge theory at non-zero temperature.

\subsection{Quark-antiquark potential in a confining background}

\begin{figure}[ht]
\center
\includegraphics[width=0.35\textwidth]{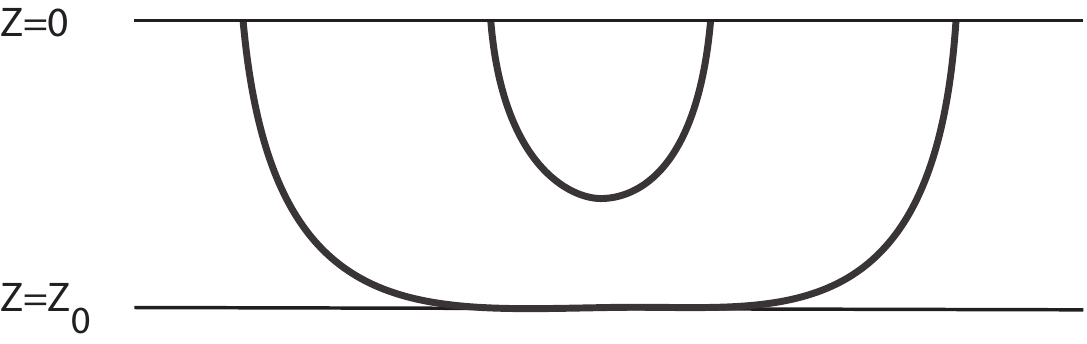}
\includegraphics[width=0.60\textwidth]{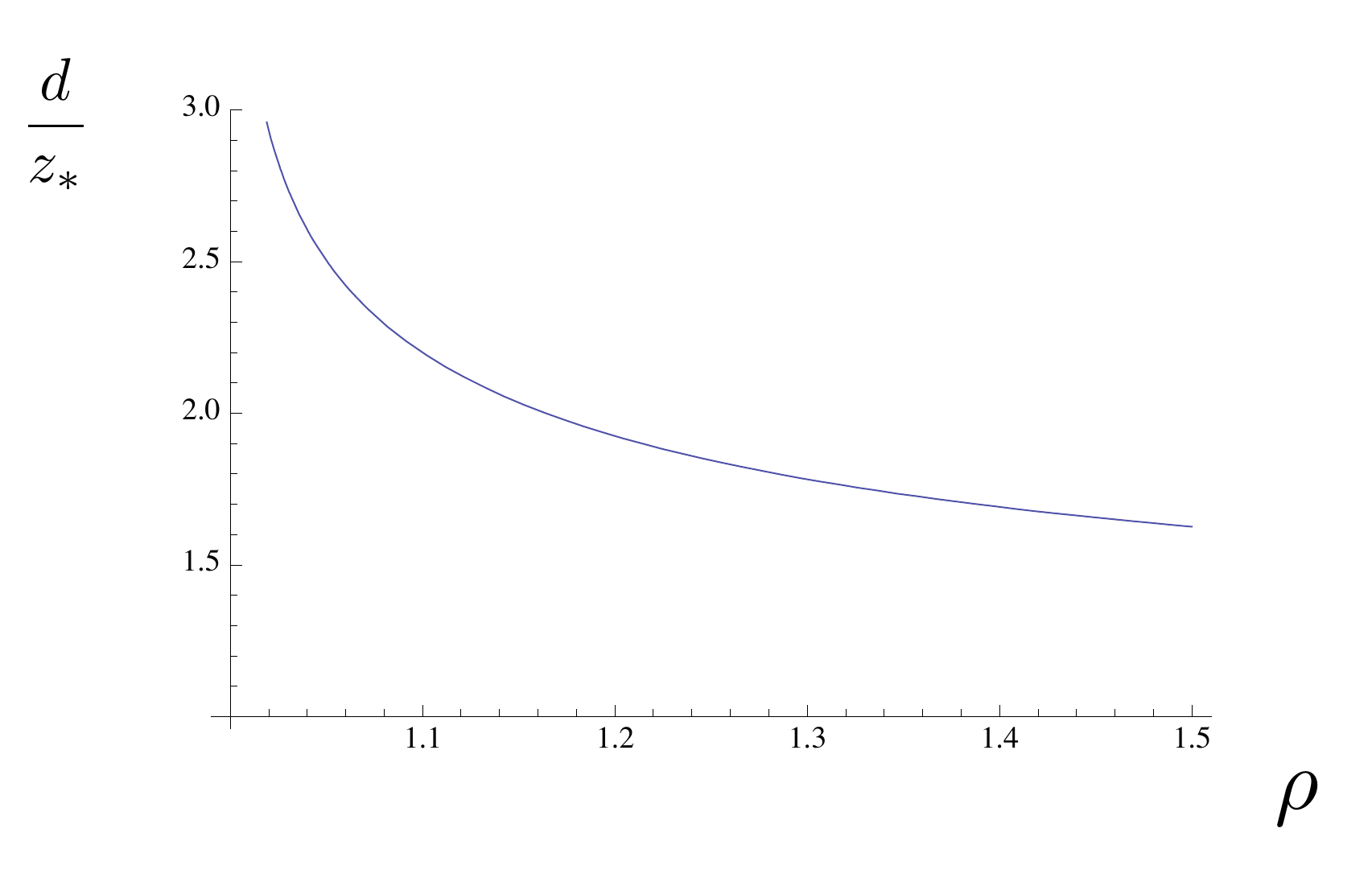}
\caption{In a confining metric the space ends smoothly and the profile of the hanging string becomes almost rectangular for large $q\bar q$ separation (left). The distance $d$  is not bounded, as shown on the plot on the right. } 
\label{Wilson_confining}
\end{figure}

Let us consider the AdS black hole in Euclidean signature and let us go back to Minkowski signature by analytic continuation along a different direction, namely by making $x_3\to it$. If we call $u$ to the original euclidean time, we arrive at the following metric:
\beq
ds^2\,=\,{L^2\over z^2}\,\Big[-\,dt^2\,+\,d x_1^{\,2}\,+\,d x_2^{\,2}\,+\,f(z) d u^2\,+\,
{dz^2\over f(z)}\,\Big]\,\,.
\label{metric_confining}
\eeq
In this case the space ends smoothly at $z=z_0$, which should be though as a IR mass scale. We will  now compute the $q\bar q$ potential for the metric (\ref{metric_confining})  and we will verify that it corresponds to a confining background. With this purpose,  let us consider a fundamental string in  the euclidean version of the above metric and let us use $(t,x)$ as worldsheet coordinates. The induced metric is:
\beq
ds^2\,=\,
{L^2\over z^2}\,\Big[\,\,dt^2+\Big(1+{z'^{\,2}\over f}\Big)\,dx^2\,\Big]\,\,,
\eeq
and the Nambu-Goto action becomes:
\beq
S\,=\,{\tau L^2\over 2\pi\alpha'}\,\int {dx \over z^2}
\sqrt{1+{z'^{\,2}\over f(z)}}\,\,,
\label{NG_action_confining}
\eeq
where $\tau=\int dt$. The first integral corresponding to the action (\ref{NG_action_confining})  is:
\beq
{z^2\over\sqrt{ f(z)}}\,\sqrt{f(z)+z'^{\,2}}\,=z_*^2\,\,,
\eeq
from which we get:
\beq
z'\,=\,\pm \sqrt{f(z)}\,\,
{\sqrt{z_*^4-z^4}\over z^2}\,\,.
\eeq
This equation can be integrated as:
\beq
x\,=\,\pm z_*\,\sqrt{\rho}\,
\int_{1}^{{z\over z_*}}\,
{y^2 dy\over \sqrt{(1-y^4)(\rho-y^4)}}\,\,,
\eeq
where the constant $\rho$ is the same as in (\ref{rho_def}). It follows that the $q\bar q$ distance $d$ is:
\beq
d\,=\,2 z_*\,\sqrt{\rho}\,
\int_{0}^{1}\,
{y^2 dy\over \sqrt{(1-y^4)(\rho-y^4)}}\,\,.
\eeq
In this case, $d$ grows without limit as $\rho\to 1$, or equivalently as the turning point approaches the end of the space (\ie\ when $z_*\to z_0$) (see figure \ref{Wilson_confining}). In the large $d$ limit the profile of the hanging string is approximately rectangular. The  energy due to the vertical parts of the profile can be identified with the masses of the static quarks, which have to be subtracted to get the potential energy.  The $q\bar q$ potential is just due to the horizontal part of the profile. Since in this part $z$ is approximately constant and equal to 
$z_0$, we get that the contribution to the euclidean action is:
\beq
S_{horizontal}\,=\,{\tau L^2\over 2\pi\alpha'}\,{d\over z_0^2}\,\,,
\eeq
which corresponds to an area law (it is proportional to $\tau d$) and gives rise to a confining potential of the type:
\beq
V\,=\,\sigma_s\,d\,\,,
\eeq
with $\sigma_s$ being the effective string tension, given by:
\beq
\sigma_s\,=\,
{L^2\over 2\pi\alpha'}\,{1\over z_0^2}\,\,.
\eeq
In terms of gauge theory quantities, $\sigma_s$ can be written as:
\beq
\sigma_s\,=\,{\sqrt{\lambda}\over 2\pi z_0^2}\,\,.
\eeq
Notice that $M\sim 1/ z_0$ is an IR scale that can be identified with  the mass gap of the theory (and $z_0$ with the glueball size). The previous formula for the tension is simply:
\beq
\sigma_s\,\sim\,\sqrt{\lambda}\,M^2\,\,.
\eeq

\section{Black hole thermodynamics}

The partition function in statistical mechanics in the canonical ensemble is given by:
\beq
Z\,=\,{\rm Tr}\,e^{-{H\over T}}\,\,,
\eeq
where, $H$ is the hamiltonian operator,  $T$ is the temperature and we are taking the Boltzmann constant $k_B=1$.  The thermal average of an operator ${\cal O}$ at the temperature $T$ is:
\beq
\langle {\cal O}\rangle_{T}={{\rm Tr}\big[{\cal O}\,e^{-{H\over T}}\big]\over Z}\,\,.
\eeq
In the path integral approach, the average $\langle {\cal O}\rangle_{T}$ can be written as:
\beq
\langle {\cal O}\rangle_{T}\sim 
\int [D\psi] \,
\langle \psi(x), t\,| {\cal O}\,e^{-{H\over T}} |\psi(x), t\rangle\,\,,
\label{VEV_OT}
\eeq
where the expectation value is taken between the same initial and final state $ |\psi(x), t\rangle$ (as it corresponds to a trace in Hilbert space).  Equivalently, since the hamiltonian operator implements time evolution,  we can rewrite 
(\ref{VEV_OT}) as:
\beq
\langle {\cal O}\rangle_{T}\sim 
\int [D\psi] \,
\langle \psi(x), t\,| {\cal O} |\psi(x), t+{i\over T}\rangle\,\,.
\label{VEV_Thermal}
\eeq
Then, to perform thermal averages we have to consider imaginary time evolution and we have to impose periodic boundary conditions in the Hilbert space (antiperiodic for fermions). Then, the euclidean time $t_E$ must be periodically identified
\beq
t_E\equiv t_E+{1\over T}\,\,.
\eeq
Thus, the  compactification of Euclidean time is equivalent to having $T\not=0$.

Let us apply these ideas to obtain the Hawking temperature of a black hole. 
We will assume that we have  an euclidean metric of the type:
\beq
ds^2\,=\,g(r)\,\big[\,f(r)\,dt_E^2\,+\,d\vec x^{\,2}\,\big]\,+\,{1\over h(r)}\,dr^2\,\,,
\label{general_BH_metric}
\eeq
where the functions $f(r)$ and $h(r)$ have a first-order zero at $r=r_0$, which is the location of the horizon, and  $g(r_0)\not=0$. Then, for $r\approx r_0$
\beq
f(r)\approx f'(r_0)\,(r-r_0)\,\,,
\qquad\qquad
h(r)\approx h'(r_0)\,(r-r_0)\,\,,
\eeq
while we can take $g(r)=g(r_0)$. Then, the near-horizon euclidean metric can be written as:
\beq
ds^2\,\approx\,g(r_0)\,\big[f'(r_0)\,(r-r_0) dt_E^2\,+\,d\vec x^{\,2}\,\big] +
{1\over h'(r_0)}\,{dr^2\over r-r_0}\,\,.
\eeq
Let us define a new radial variable $\rho$ such that:
\beq
{1\over h'(r_0)}\,{dr^2\over r-r_0}\,=\,d\rho^2\,\,,
\eeq
which can be integrated to give the following relation between $\rho$ and $r$:
\beq
\rho\,=\,2\,\sqrt{{r-r_0\over h'(r_0)}}\,\,.
\eeq
Next, we define an angular coordinate $\theta$ such that
\beq
g(r_0)\,f'(r_0)\,(r-r_0) dt_E^2\,=\,\rho^2\,d\theta^2\,\,,
\eeq
which is equivalent to defining $\theta$ as:
\beq
\theta={1\over 2}
\sqrt{g(r_0)\,f'(r_0)\,h'(r_0)}\,\,t_E\,\,.
\eeq
In the new variables, the $(t_E,r)$ part of the metric takes the form $d\rho^2+\rho^2 d\theta^2$, which is locally like the flat metric of a plane. In order to have $\rho=0$ (\ie\ the horizon) as a regular point without any curvature singularity, the variable  $\theta$ must be a periodic variable with period $2\pi$.  Otherwise we would have a conical singularity at the origin, due to the defect angle. We have argued  above  for a general system that the compactification of the euclidean time is equivalent to having a non-zero temperature $T$. Thus, it follows that we can assign a temperature $T$ to a black hole (the Hawking temperature). In order to find the value of $T$, let us notice that the periodicity under $\theta\to \theta+2\pi$ is equivalent to periodicity under $t_E\to t_E+{1\over T}$, where $T$ is the Hawking temperature given by:
\beq
{1\over T}\,=\,{4\pi\over \sqrt{g(r_0)\,f'(r_0)\,h'(r_0)}}\,\,.
\eeq
In the next subsections we will apply this formula to determine the Hawking temperature of a couple of black holes.

\subsection{Application to the Schwarzschild black hole}

Let us consider the  ordinary Schwarzschild  metric in four dimensions:
\beq
ds^2\,=\,-\Big(1-{2GM\over r}\Big)\,dt^2\,+\,{dr^2\over 1-{2GM\over r}}\,+\,
r^2 d\Omega_2^2\,\,,
\label{Sch_BH}
\eeq
which is a particular case  of the general  expression written in (\ref{general_BH_metric}),  with the functions $f$, $h$ and $g$ being given by:
\beq
g(r)=1\,\,,
\qquad\qquad
f(r)=h(r)=1-{2GM\over r}\,\,.
\eeq
In (\ref{Sch_BH}) $G$ is the four-dimensional Newton constant and $M$ is the mass of the black hole. The horizon in the geometry (\ref{Sch_BH}) is located at $r=r_0=2 GM$. Since:
\beq
f'(r_0)=h'(r_0)={2GM\over r_0^2}
\eeq
it follows that the Hawking temperature for the  Schwarzschild black hole is given by:
\beq
T\,=\,{1\over 8\pi G M}\,\,.
\label{Hawking_T_Sch}
\eeq
Let us use the expression of $T$  in (\ref{Hawking_T_Sch}) to obtain the black hole entropy $S$ of the Schwarzschild black hole. We will identify the mass $M$ with the internal energy and we  will make use of the first law of thermodynamics:
\beq
dM=T dS={1\over  8\pi G M} dS\,\,,
\eeq
which can be integrated to give  the entropy $S$:
\beq
S\,=\,4\pi\,G\,M^2\,\,,
\eeq
The horizon area $A_H$  for the metric (\ref{Sch_BH}) is the area of the surface $t={\rm constant}$, $r=r_0$. It is straightforward to demonstrate that $A_H$ is:
\beq
A_H=4\pi r_0^2\,=\,16\,G^2\,M^2\,\,.
\eeq
It is now immediate to check that the relation between $S$ and $A_H$ is:
\beq
S\,=\,{A_H\over 4 G}\,\,,
\eeq
which is noting but the celebrated Bekenstein-Hawking entropy formula which relates the entropy of a black hole with the area of its horizon (see (\ref{BH-formula})). 

\subsection {AdS black hole}
\label{section_AdS_BH}

The metric of a black hole in $AdS_{d+1}$ is:
\beq
ds^2\,=\,{L^2\over z^2}\,\Big[\,f(z)\,dt_E^2\,+\,d\vec x^{\,2}\,+\,{dz^2\over f(z)}\,\Big]\,\,,
\label{AdS_BH_metric_in_z}
\eeq
 where $f(z)$ is the following function:
 \beq
 f(z)\,=\,1\,-\,{z^d\over z_0^d}\,\,,
 \eeq
with $z_0$ being a constant ($z=z_0$ is the location of the horizon). 
This metric is  just of our general form (\ref{general_BH_metric}) with $r\to z$, $f$ given as above and $g$  and $h$ being:
 \beq
 g={L^2\over z^2}
 \,\,, \qquad\qquad
 h\,=\,{z^2\over L^2}\,f\,\,.
 \eeq
In this case the derivatives of $f$ and $h$ at the horizon are:
\beq
f'(z_0)=-{d\over z_0}
 \,\,, \qquad\qquad
 h'(z_0)=-{d z_0\over L^2}\,\,,
 \eeq
and, therefore, we have:
\beq
g(z_0)\,f'(z_0)\,h'(z_0)={d^2\over z_0^2}\,\,.
\eeq
Thus, the Hawking temperature is related to $z_0$ by means of the relation:
\beq
T\,=\,{d\over 4\pi z_0}\,\,.
\eeq
The horizon is the hypersurface $z=z_0$ and $t$ constant, whose area  in the metric (\ref{AdS_BH_metric_in_z}) is:
\beq
A_H\,=\,\Big({L\over z_0}\Big)^{d-1}\,V_{d-1}\,\,,
\eeq
where $V_{d-1}$ is the volume along the Minkowski spacial directions. In terms of the temperature, $A_H$ can be written as:
\beq
A_H\,=\,\Big({4\pi\over d}\Big)^{d-1}\,L^{d-1}\,T^{d-1}\,
V_{d-1}\,\,,
\eeq
and the entropy can be computed from the Bekenstein-Hawking formula, with the result:
\beq
S\,=\,{A_H\over 4G_{d+1}}\,=\,{1\over 4G_{d+1}}\,
\Big({4\pi\over d}\Big)^{d-1}\,L^{d-1}\,T^{d-1}\,
V_{d-1}\,\,. 
\eeq
We now define the entropy density $s$ as:
\beq
s\,=\,{S\over V_{d-1}}\,\,.
\eeq
Let us write $s$ in terms of the QFT central charge $c_{QFT}={1\over 4}
(L/l_P)^{d-1}$. We get:
\beq
s\,=\,\Big({4\pi\over d}\Big)^{d-1}\,c_{QFT}\, T^{d-1}\,\,.
\eeq
In the case of  ${\cal N}=4$ SYM, we take  $d=4$ and $c_{SYM}= N^2/2\pi$ and the entropy density is given by:
\beq
s_{SYM}={\pi^2\over 2}\,N^2\,T^3\,\,.
\label{entropy_density_YM}
\eeq
From the expression of the entropy density  in (\ref{entropy_density_YM}) we can obtain the value of the pressure by means of the thermodynamic relation:
\beq
s\,=\,{\partial p\over \partial T}\,\,.
\label{s_p}
\eeq
We get:
\beq
p\,=\,{\pi^2\over 8}\,N^2\,T^4\,\,.
\eeq
Moreover, the energy density $\epsilon$ can be obtained from $p$ and $s$ by means of the standard thermodynamic relation:
\beq
\epsilon\,=\,-p+Ts\,\,.
\label{epsilon_p_s}
\eeq
Using the values of $p$ and $s$ computed from holography, we arrive at the following value of the energy density:
\beq
\epsilon\,=\,{3\pi^2\over 8}\,N^2\,T^4\,\,.
\eeq

\subsection{ Comparison with field theory}
Let us compare the strong coupling values of  $s$, $p$ and $\epsilon$  found above with those corresponding the ${\cal N}=4$ SYM at zero coupling.  The partition function in the canonical ensemble for  a gas of  non-interacting relativistic bosons and fermions is 
\beq
\log Z=\mp V_3\,\int {d^3p\over (2\pi)^3}\,\,\log \Big(
1\mp e^{-{\omega(p)\over T}}\Big)\,\,,
\eeq
where  the upper minus (lower plus) signs corresponds to bosons (respectively, fermions) and $\omega(p)=\sqrt{\vec p^{\,2}+m^2}$. In the massless case, we just take $\omega(p)=|\vec p|$ and we get:
\beq
{\log Z\over V_3}\,=\,\mp \int_0^{\infty}\,{dp\over 2\pi^2}\,p^2\,\log\Big(1\mp e^{-{p\over T}}\Big)\,\,.
\label{logZ_massless}
\eeq
Let us change variables in the integral (\ref{logZ_massless}) from $p$ to $x=p/T$. After integrating by parts we find:
\beq
{\log Z\over V_3}\,=\,{T^3\over 6\pi^2}\,\int_{0}^{\infty}\,dx\,\,
{x^3\over e^x\mp 1}\,\,.
\eeq
To calculate these integrals we use the general results, valid for $n\in {\mathbb Z}$:
\beq
\int_{0}^{\infty}\,dx\,\,{x^{2n-1}\over e^x+1}\,=\,{2^{2n-1}-1\over 2n}\,\pi^{2n}\,B_n
\,\,,
\qquad\qquad
\int_{0}^{\infty}\,dx\,\,{x^{2n-1}\over e^x-1}\,=\,{(2\pi)^{2n}\,B_n\over 4n}\,\,,
\eeq
where $B_n$ denotes the Bernouilli numbers. In particular for $n=2$, since 
$B_2\,=\,1/ 30$, we have:
\beq
\int_{0}^{\infty}\,dx\,\,{x^{3}\over e^x-1}\,=\,{\pi^4\over 15}
\,\,,
\qquad\qquad
\int_{0}^{\infty}\,dx\,\,{x^{3}\over e^x+1}\,=\,{7\pi^4\over 120}\,\,.
\eeq
Thus, for bosons:
\beq
{\log Z\over V_3}\,=\,{\pi^2\over 90}\,T^3\,\,,
\qquad\qquad
({\rm bosons})\,\,,
\eeq
while for fermions:
\beq
{\log Z\over V_3}\,=\,{7\pi^2\over 720}\,T^3\,\,,
\qquad\qquad
({\rm fermions})\,\,.
\eeq
The entropy density can be obtained from the partition function by means of the standard 
statistical mechanics relation:
\beq
s\,=\,{\partial \over \partial T}\,\Big[T\,{\log Z\over V_3}\Big]\,=\,4\,{\log Z\over V_3}\,\,.
\eeq
Then, it follows that:
\beq
s_{boson}\,=\,{2\pi^2\over 45}\,T^3
\,\,,
\qquad\qquad
s_{fermion}\,=\,{7\pi^2\over 180}\,T^3\,\,.
\eeq
In ${\cal N}=4$ SYM the number of bosons we have is:
\beq
\big[\,2({\rm gauge \,field})+6 ({\rm scalar \,field})\,\big]\,N^2\,=\,8N^2\,\,,
\eeq
while the number of fermions is:
\beq
\big[\,2\times 4({\rm Weyl\,spinors})\,\big]\,N^2\,=\,8N^2\,\,,
\eeq
which, due to the supersymmetry, is the same as the number of bosons. Therefore, the total entropy  density  of  ${\cal N}=4$ SYM  when the coupling constant is zero is:
\beq
s_{{\cal N}=4\,free\, gas}\,=\,
8\,N^2\,\Big[{2\pi^2\over 45}\,+\,{7\pi^2\over 180}\,\Big]\,T^3\,=\,{2\pi^2\over 3}\,
N^2\,T^3\,\,.
\eeq
By using the thermodynamic formulas (\ref{epsilon_p_s}) and (\ref{s_p}), we obtain the values of the pressure and energy density:
\beq
p_{{\cal N}=4\,free\, gas}\,=\,{\pi^2\over 6}\,
N^2\,T^4\,\,,
\qquad\qquad
\epsilon_{{\cal N}=4\,free\, gas}\,=\,{\pi^2\over 2}\,
N^2\,T^4\,\,.
\eeq
Then, by comparing with the result given by the black hole, we obtain:
\beq
s_{black\, hole}\,=\,{3\over 4}\,s_{{\cal N}=4\,free\, gas}\,\,,
\label{sbh-sgass}
\eeq
and similarly for the pressure and energy density:
\beq
p_{black\, hole}\,=\,{3\over 4}\,p_{{\cal N}=4\,free\, gas}\,\,,
\qquad\qquad
\epsilon_{black\, hole}\,=\,{3\over 4}\,\epsilon_{{\cal N}=4\,free\, gas}\,\,.
\label{p-epsilon_bh-sgass}
\eeq
Therefore, the AdS/CFT correspondence predicts that the values of $s$, $p$ and $\epsilon$ at infinite coupling differ from their values at   zero coupling by a multiplicative  factor $3/4$. These reductions of the entropy, pressure and energy density as the coupling is increased are in very good agreement with the results obtained in lattice simulations for different gauge theories.

\section{Transport coefficients}
\label{transport_coefficients}

Let us  begin by writing the linear response formulas of section \ref{Linear_response} in real time. We consider a system in QFT to which we couple a source $\varphi(x)$ to a local operator ${\cal O}(x)$:
\beq
S\,=\,S_0\,+\,\int d^dx \,{\cal O}(x)\,\varphi(x)\,\,,
\eeq
where the $d$-dimensional spacetime has Minkowski signature.  We will assume that the unperturbed VEV of the operator ${\cal O}$ vanishes. Then, the one-point function of the operator ${\cal O}$ in the presence of the source is given by:
\beq
\langle {\cal O} (x)\rangle_{\varphi} \,=\,-\int G_{R} (x-y)\,\varphi(y)\,dy\,\,,
\label{VEV_linear_response_real}
\eeq
where $G_{R} (x-y)$ is the retarded Green's function, defined as:
\beq
i\,G_{R} (x-y)\,\equiv\,\theta(x^0-y^0)\,\langle \big[
{\cal O}(x), {\cal O}(y)\big]\rangle\,\,.
\eeq
The fact that the  linear response is determined by the retarded correlator is a consequence of causality since  the source can influence the system only after it has been turned on. In momentum space the relation (\ref{VEV_linear_response_real}) between  the one-point function and the source becomes:
\beq
\langle {\cal O} (\omega, \vec k)\rangle_{\varphi} =
-G_R(\,\omega, \vec k\,) \,\varphi (\,\omega, \vec k\,) \,\,.
\label{VEV_lin_res_real_k}
\eeq
We are interested in analyzing the long wavelength hydrodynamic limit. In this case one can take the zero spatial momentum and zero frequency limit of the retarded correlator. 
In this limit one studies the response of the system to a time varying source $\varphi(t)$, which can be approximately represented as:
\beq
\langle {\cal O} \rangle_{\varphi} \approx -\chi\,\, \partial_t\,\varphi\,\,,
\label{VEV_O_chi}
\eeq
where  the real constant $\chi$ is the so-called transport coefficient. In frequency space, for $\omega\to 0$,  the relation  (\ref{VEV_O_chi}) is given by:
\beq
\langle {\cal O} \rangle_{\varphi} \approx  i \,\omega\, \chi\,\varphi (\omega)\,\,.
\label{VEV_O_chi_k}
\eeq
The linear response equation for this quantity is obtained from the long wavelength limit $\vec k\to 0$ of (\ref{VEV_lin_res_real_k}):
\beq
\langle {\cal O}\rangle_{\varphi} =
-G_R(\,\omega, \vec k =0\,) \,\varphi (\,\omega)\,\,.
\label{linear_response_hydro}
\eeq
By comparing (\ref{VEV_O_chi_k}) and (\ref{linear_response_hydro}) we get:
\beq
G_R(\omega, \vec k=0)\,=\,-i\omega\,\chi\,\,,
\qquad\qquad (\omega\to 0)\,\,,
\eeq
or, taking the imaginary part:
\beq
{\rm Im}\, G_R(\omega, \vec k=0)\,=\,-\omega\,\chi\,\,,
\qquad\qquad (\omega\to 0)\,\,.
\label{ImG_R}
\eeq
From (\ref{ImG_R}) we immediately get the so-called  Kubo formula for $\chi$:
\beq
\chi\,=\,-\lim_{\omega\to 0}\,\lim_{\vec k\to 0}\, {1\over \omega}\,
{\rm Im}\, G_R(\omega, \vec k)\,\,.
\label{Kubo}
\eeq

Let us now see how we can make use of (\ref{Kubo}) to compute the transport coefficient $\chi$ by using holographic methods.  Let us consider a $(d+1)$-dimensional metric of the form:
\beq
ds^2\,=\,g_{tt}\,dt^2+g_{zz} dz^2+g_{xx} \delta_{ij}\,dx^i dx^j\,\,.
\label{metric_for_trasnport}
\eeq
We will  assume the metric (\ref{metric_for_trasnport})  has an horizon at $z=z_0$ and that  $g_{tt}$ and that $g_{zz}$ behave near $z=z_0$ as:
\beq
g_{tt}\approx -c_0 (z_0-z)
\,\,\,,
\qquad
g_{zz}\approx {c_z \over z_0-z}\,\,,
\qquad\qquad
z\to z_0\,\,,
\eeq
with $c_0$ and $c_z$ being constants. Let us consider a massless scalar field $\phi$ in this metric such that its action is:
\beq
S\,=\,-{1\over 2}\,\int\,d^{d+1}\,x\,\sqrt{-g}\,{\partial_M\,\phi\,\partial^M\,\phi\over q(z)}\,\,,
\label{S_transport}
\eeq
where the function $q(z)$ is an effective coupling of the mode. The Euler-Lagrange equation of motion  derived from (\ref{S_transport}) is:
\beq
\partial_{M}\,\Big({\sqrt{-g}\over q}\,g^{MN}\,\partial_M\,\phi\,\Big)\,=\,0\,\,.
\label{eom_trasnport}
\eeq
In terms of the canonical  momentum $\Pi$, defined as (see (\ref{Pi})):
\beq
\Pi\equiv- {\partial {\cal L}\over \partial(\partial_z \phi)}\,=\,
{\sqrt{-g}\over q}\,g^{zz}\,\partial_z\,\phi\,\,,
\label{Pi_transport}
\eeq
the equation of motion  (\ref{eom_trasnport}) becomes:
\beq
\partial_z\,\Pi=-{\sqrt{-g}\over q}\,\Big(\,
{\partial_t^{2}\phi\over g_{tt}}+{\partial_i^{2}\phi\over g_{xx}}
\,\Big)\,\,.
\eeq
Let us now Fourier transform $\phi$ and $\Pi$  in the variables $t, \vec x$:
\bear
&&\phi(z,t,\vec x)\,=\,\int {d\omega\, d^{d-1} k\over (2\pi)^d}\,
e^{i(\vec k\cdot x-\omega t)}\, \phi (z,\omega, \vec k)\,\,,
\rc\rc
&&\Pi(z,t,\vec x)\,=\,\int {d\omega\, d^{d-1} k\over (2\pi)^d}\,
e^{i(\vec k\cdot x-\omega t)}\, \Pi (z,\omega, \vec k)\,\,.
\eear
Since in this massless case the scaling dimension $\Delta$ is equal to $d$ (see (\ref{dimension_mass}) for $m=0$), the one-point function is just the limit of the momentum $\Pi$ at the boundary $z=0$ and the boundary field $\varphi$  is just obtained by taking the limit $z\to 0$ of the bulk field $\phi$, without any multiplicative factor (see  eqs. (\ref{one-point-Pi}) and (\ref{renor-field})). It follows from (\ref{VEV_O_chi_k}) that the transport coefficient $\chi$ is obtained as:
\beq
\chi\,=\,\lim_{k_{\mu}\to 0}\,\lim_{z\to 0} {\rm Im}\,
\Bigg[{\Pi(z,k_{\mu})\over \omega\, \phi(z,k_{\mu})}\,\Bigg]\,=\,
\lim_{k_{\mu}\to 0}\,\lim_{z\to 0} \,
{\Pi(z,k_{\mu})\over i \omega\, \phi(z,k_{\mu})}\,\,,
\eeq
where, in the last step, we used the fact that ${\rm Re}\Pi/\omega\to 0$ as $\omega\to 0$
(see below). It turns out that evaluating $\Pi/ (\omega\, \phi)$ at the boundary is equivalent  to evaluate it at the horizon. In order to prove this,  let us consider the general equation:
\beq
\partial_z\,\big[\,A(z)\,\partial_z\phi\,\big]\,=\,B(z)\,\phi(z)\,\,.
\label{general_eq}
\eeq
Let us write (\ref{general_eq})  in hamiltonian form. We first define:
\beq
P(z)\,\equiv A(z)\,\partial_z \phi(z)\,\,.
\eeq
Then, (\ref{general_eq})  can be written as:
\beq
\partial_z\,P(z)\,=\,B(z)\,\phi(z)\,\,.
\label{general_eq_Pi}
\eeq
Then, one can readily prove that (\ref{general_eq}) and (\ref{general_eq_Pi}) can be combined  in the following first-order  Riccati equation:
\beq
\partial_z\,\Big({P(z)\over \phi(z)}\Big)\,=\,B(z)\,-{1\over A(z)}\,
\Big({P(z)\over \phi(z)}\Big)^2\,\,.
\eeq
For the equation of motion  of the scalar field $\phi$ in momentum space,
the functions $A(z)$ and $B(z)$ are given by:
\beq
A(z)={\sqrt{-g}\over q}\,g^{zz}\,\,,
\qquad\qquad
B(z)={\sqrt{-g}\over q}\,\Big[{\omega^2\over g_{tt}}\,+\,{{\vec k}^{\,2}\over g_{xx}}\,
\Big]\,\,,
\eeq
and $P(z)=\Pi(z)$ (see (\ref{Pi_transport})). 
Then, we can write:
\beq
\partial_z\,\Bigg[\,{\Pi\over \omega \phi}\,\Bigg]\,=\,-
\omega\Bigg[ {q g_{zz}\over \sqrt{-g}}\,
\Bigg(\,{\Pi\over \omega \phi}\,\Bigg)^2\,-\,
{\sqrt{-g}\over q\,g_{tt}}\,\Big(1\,+\,{g_{tt}\over g_{xx}}\,\,
{\vec k^{\,2}\over \omega^2}\,\Big)\,\Bigg]\,\,.
\label{holographic_evolution}
\eeq
The right-hand side of equation (\ref{holographic_evolution}) vanishes when the ordered limit  $\lim_{\omega \to 0}\,\lim_{\vec k\to 0}$ is taken. It follows that 
$\Pi/ (\omega \phi)$ is independent of $z$ in this limit and, as claimed above, it can be evaluated at the horizon. Thus, we can write the transport coefficient $\chi$ as:
\beq
\chi\,=\,
\lim_{k_{\mu}\to 0}\,\lim_{z\to z_0} \,
{\Pi(z,k_{\mu})\over i \omega\, \phi(z,k_{\mu})}\,\,.
\label{chi_horizon}
\eeq
In order to evaluate the right-hand side of (\ref{chi_horizon}), let us study the equation of motion near  $z=z_0$, where $\Pi$ becomes:
\beq
\Pi \approx {1\over c_z}\, {\sqrt{-g(z_0)}\over q(z_0)}\,
(z_0-z)\,\partial_z\, \phi\,\,,
\eeq
and the equation of motion takes the form:
\beq
\partial_z\,\Big[\,(z_0-z)\,\partial_z\,\phi(z, k_{\mu}) \,\Big]\,+\,
 c_z\,\Big[\,{\omega^2\over c_0 (z_0-z)}\,-\,
 {k^2\over g_{xx}(z_0)}\,\Big]\, \phi(z, k_{\mu}) \,=\,0\,\,.
 \label{near_horizon_phi}
 \eeq
To find an approximate solution of (\ref{near_horizon_phi}) near the horizon $z=z_0$, we neglect the last term in the previous equation and try  to find a solution of the type:
\beq
\phi = (z_0-z)^{\beta}\,\,.
\label{near_horizon_ansatz}
\eeq
Plugging the ansatz (\ref{near_horizon_ansatz}) in (\ref{near_horizon_phi}),  we get that the exponent $\beta$ can  take the following two values:
\beq
\beta=\pm i\,\sqrt{{c_z\over c_0}}\,\,\omega\,\,,
\eeq
which correspond to the following two solutions:
\beq
\phi_{\pm}\sim 
(z_0-z)^{\pm i\,\sqrt{{c_z\over c_0}}\,\,\omega}\,\,.
\label{phi_pm}
\eeq
Only one of the two solutions in (\ref{phi_pm}) is compatible with causality. Indeed, let us define a new variable $r$ as $z_0-z=e^r$. In the $r$ variable the horizon is located at $r\to-\infty$. By inserting the $t$ dependence, the two solutions $\phi_{\pm}$ are:
\beq
\phi_{\pm}\sim e^{-i(\omega \,t\,\mp\,\Omega\,r)}\,\,,
\label{phi_pm_r}
\eeq
where $\Omega=\omega\sqrt{c_z/ c_0}$. Clearly, $\phi_-$ is an incoming wave at the horizon since if we increase $t\to t+\epsilon$, we must decrease $r$ as $r\to r-\epsilon\,\Omega/\omega$ to keep $\phi_-$ constant. Then, the wave  $\phi_-$  moves towards  the horizon $r\to-\infty$ and is an  infalling wave at the horizon. Similarly, $\phi_{+}$  is an outgoing wave at the horizon.  Causality on the gravity side, is implemented if 
we choose our solution to be  the infalling one $\phi_{-}$. One can show that this corresponds to having retarded Green's functions on the field theory side. This  infalling solution near $z=z_0$ satisfies:
\beq
\partial_z \phi_{-}=
\sqrt{{g_{zz}(z_0)\over -g_{tt}(z_0)}}\,\,i\omega\, \phi_{-}\,\,,
\eeq
and, therefore, one has:
\beq
{\Pi\over i\omega\, \phi_-}\Bigg|_{z_0}\,=\,
{1\over q(z_0)}\,
\sqrt{{g\over g_{zz}\,g_{tt}}}\Bigg|_{z_0}\,\,.
\eeq
It follows that the transport coefficient $\chi$ is given by:
\beq
\chi={1\over q(z_0)}\,
\sqrt{{g\over g_{zz}\,g_{tt}}}\Bigg|_{z_0}\,\,.
\eeq
Notice that the square root in this last equation is just the area of the horizon $A_H$ divided by the spatial volume $V$. Then, we can alternatively write the transport coefficient $\chi$ as:
\beq
\chi={1\over q(z_0)}\,{A_H\over V}\,\,.
\label{chi_general_scalar}
\eeq
In particular, for an AdS black hole, the previous expression becomes:
\beq
\chi={1\over q(z_0)}\,\Big({L\over z_0}\Big)^{d-1}\,\,,
\qquad\qquad
({\rm AdS\,\,black\,\,hole})\,\,.
\eeq
More interestingly, one can compare the general formula (\ref{chi_general_scalar}) with the entropy density, as given by the Bekenstein-Hawking formula $s=A_{H}/ (4G_N V)$. By computing the ratio $\chi/s$, we get the simple result: 
\beq
{\chi\over s}={4G_N\over  q(z_0)}\,\,.
\eeq

\section{Holographic viscosities}
Hydrodynamics can be thought as an effective theory describing the dynamics of a continuous system at large distances and time scales. In order to study the dynamics of the system we have to analyze the energy-momentum tensor $T^{\mu\nu}$ which  is conserved ($\partial_{\mu}T^{\mu\nu}=0$). We will assume that the system is at local thermal equilibrium and that the state of the system at a given time is determined by the local temperature $T(x)$ and the local fluid velocity $u^{\mu}(x)$, which satisfies the condition $u_{\mu}u^{\mu}=-1$. 

The relation between $T^{\mu\nu}$ and $u^{\mu}$ and the thermodynamic functions is expressed by means of the so-called constitutive relations, which for isotropic fluids takes the form:
\beq
T^{\mu\nu}\,=\,(\epsilon+p)\,u^{\mu} u^{\nu}\,+\,p\,g^{\mu\nu}\,-\,\sigma^{\mu\nu}\,\,,
\eeq
where $\epsilon$ is the energy density and $p$ is the pressure, while $\sigma^{\mu\nu}$ is the so-called dissipative part of $T^{\mu\nu}$ and depends on the derivatives of $T(x)$ and $u^{\mu}$.  In order to parametrize $\sigma_{\mu\nu}$ at first-order in the derivatives, let us consider a local rest frame in which $u^{i}(x)=0$ (and $u^{\mu}=(1,0,0,0)$). One can choose this frame in such a way that the dissipative corrections to the energy-momentum tensor components $T^{0\mu}$ vanish. Then, $\sigma^{00}=\sigma^{0i}=0$ or, equivalently, $T^{00}=\epsilon$,  $T^{0i}=0$. The only non-zero elements of the dissipative energy-momentum  tensor are $\sigma_{ij}$. At first order in derivatives $\sigma_{ij}$ can be written as:
\beq
\sigma_{ij}\,=\,\eta \big(\partial_{i}u_j\,+\,\partial_{j} u_i\,-\,{2\over 3}\,\delta_{ij}\,\partial_k u_k\big)\,+\,\zeta\,\delta_{ij} \partial_k u_k\,\,,
\eeq
where $\eta$ is the so-called shear viscosity and $\zeta$ is the bulk viscosity.

Let us write $\sigma^{\mu\nu}$ in covariant form. We first define the projector onto the directions perpendicular to $u^{\mu}$, as:
\beq
P^{\mu\nu}\,=\,g^{\mu\nu}\,+u^{\mu} u^{\nu}\,\,.
\eeq
Then, in an arbitrary curved metric $\sigma_{\mu\nu}$ can be written as:
\beq
\sigma^{\mu\nu}\,=\,P^{\mu\alpha}\,P^{\nu\beta}\,\Big[\,
\eta \big(\nabla_{\alpha} u_{\beta}+\nabla_{\beta}u_{\alpha}\big)\,+\,
\Big(\zeta-{2\over 3}\eta\Big)\,g_{\alpha\beta}\,\nabla\cdot u\,\Big]\,\,,
\eeq
where the covariant derivatives  of the vectors $u_{\beta}$ are defined as:
\beq
\nabla_{\alpha} u_{\beta}\,=\,\partial_{\alpha} u_{\beta}\,-\,
\Gamma_{\alpha\beta}^{\mu}\,u_{\mu}\,\,,
\eeq
with $\Gamma_{\alpha\beta}^{\mu}$ being the Christoffel symbols, defined as:
\beq
\Gamma_{\alpha\beta}^{\mu}\,=\,{1\over 2}\,g^{\mu\lambda}\,
\Big[\,{\partial g_{\lambda\alpha}\over \partial x^{\beta}}\,+\,
{\partial g_{\lambda\beta}\over \partial x^{\alpha}}\,-\,
{\partial g_{\alpha\beta}\over \partial x^{\lambda}}\,\Big]\,\,.
\eeq
We will find $\sigma^{\mu\nu}$ as  (minus) the one-point function of $T^{\mu\nu}$ in the presence of a metric perturbation of the type $g^{\mu\nu}\to \eta^{\mu\nu}+h^{\mu\nu}$. By using linear response theory (eq.  (\ref{VEV_linear_response_real})),  we obtain:
\beq
\sigma^{\mu\nu}(x)\,=\,\int G_{R}^{\mu\nu, \alpha\beta} (x-y)\,h_{\alpha\beta} (y) dy\,\,,
\eeq
where the retarded correlator is just:
\beq
i\,G_{R}^{\mu\nu, \alpha\beta} (x-y)=\theta(x^0-y^0)\,
\langle \big[T^{\mu\nu}(x), T^{\alpha\beta}(y)\big]\rangle\,\,.
\eeq
Let us now  consider the following metric perturbation:
\beq
g_{00}(t,\vec x)=-1\,\,,
\qquad
g_{0i}(t,\vec x)=0\,\,,
\qquad
g_{ij}(t,\vec x)=\delta_{ij}+h_{ij}(t)\,\,.
\eeq
with $h_{ij}<<1$ and such that is traceless ($h_{ii}=0$). We will assume that 
$h_{ij}$ is a function of $t$ and that this variation with $t$ is slow.
The inverse metric is just:
\beq
g^{00}(t,\vec x)=-1\,\,,
\qquad
g^{0i}(t,\vec x)=0\,\,,
\qquad
g^{ij}(t,\vec x)=\delta_{ij}-h_{ij}(t)\,\,,
\eeq
from which we get the different components of the projector $P^{\mu\,\nu}$:
\beq
P^{00}=0\,\,,
\qquad
P^{0i}=0\,\,,
\qquad
P^{ij}=\delta_{ij}\,-\,h_{ij}\,\,.
\eeq
At first order in the perturbation, the Christoffel symbols are given by:
\beq
\Gamma_{00}^{0}\,=\,\Gamma_{0i}^{0}\,=\,0\,\,,
\qquad\qquad
\Gamma_{ij}^{0}\,=\,{1\over 2}\,\partial_{0}\, h_{ij}\,\,.
\eeq
Therefore the covariant derivatives of the velocity are:
\beq
\nabla_{0} u_{0}\,=\,\nabla_{0} u_{i}\,=\,0\,\,,
\qquad\qquad
\nabla_{i} u_{j}\,=\,{1\over 2}\,\partial_{0}\, h_{ij}\,\,.
\eeq
From these values and the hypothesis that the metric perturbation is traceless it follows that the covariant divergence of the velocity vanishes:
\beq
\nabla\cdot u={1\over 2}\,\partial_0\,h_{ii}\,=\,0\,\,.
\eeq
Let us assume that the only non-zero value of $h_{ij}$ is $h_{12}$. Then, the linear response value of $\sigma^{12}$ in frequency space is:
\beq
\sigma^{12}(\omega)=G_{R}^{12,12} (\omega, \vec k=0)\,h_{12}(\omega)\,\,.
\eeq
Moreover, by using the values of the covariant derivatives of the velocity we get 
\beq
\sigma^{12}(t)\,=\,\eta\,\partial_0 h_{12}(t)\,\,,
\eeq
or,  in frequency space (for low $\omega$):
\beq
\sigma^{12}(\omega)=-i\eta\,\omega \,h_{12}(\omega)\,\,.
\eeq
By comparing these two expressions of $\sigma^{12}$, we obtain again Kubo formula for the shear viscosity, namely:
\beq
\eta\,=\,-\lim_{\omega\to 0}\,\Big[{1\over \omega}\, {\rm Im} \,
G_{R}^{12,12} (\omega, \vec k=0)\,\Big]\,\,.
\eeq

In order to compute holographically the retarded correlator of $T_{12}$, let us consider a general $d+1$-dimensional diagonal metric and let us perturb it by adding a non-diagonal element along $x^1\, x^2$:
\beq
ds^2\,=\,g_{tt}\,dt^2+g_{zz} dz^2+g_{xx}\big( \delta_{ij}\,dx^i dx^j+2\phi\, dx^1 dx^2\big)\,\,,
\eeq
where $\phi$ is small and independent of $x^1$ and $x^2$. Notice that, in the perturbed metric at first order in $\phi$, one has:
\beq
g_{12}=g_{xx}\,\phi\,\,,
\qquad
g^{1}_{\,\,\,2}=\phi\,\,.
\eeq
Let us write the $x^1 x^2$ part of the metric (at first order) as:
\beq
g_{xx} (dx^1)^2+ g_{xx}(dx^2+\phi\, dx^1)^2\,\,,
\eeq
which is clearly similar to the ansatz corresponding to a Kaluza-Klein reduction along $x^2$ with KK gauge field connection:
\beq
A=\phi\,dx^1\,\,.
\eeq
Therefore, we can use the known results of the KK reduction to write the  quadratic action for $\phi$. Indeed, the Einstein-Hilbert action for the metric leads to the following 
expression for the action of the gauge connection $A$ (or equivalently the metric perturbation $\phi$):
\beq
S_{\phi}\,=\,-{1\over 16\pi G_N}\,
\int d^{d+1} x\, \sqrt{-g}\, g_{xx}\, {F^2\over 4}\,\,,
\eeq
where $g$ is the determinant of the unperturbed metric (with $\phi=0$), $F=dA$ is the field strength of  $A$ and in  $F^2=F_{\mu\nu}\,F^{\mu\nu}$ the indices are raised with the unperturbed  metric.  As:
\beq
F\,=\,\partial_t\phi\,dt\wedge dx^1\,+\,\partial_3\phi\, dx^3\wedge dx^1\,+\,
\partial_z\phi\, dz\wedge dx^1\,\,,
\eeq
we have:
\beq
F^2\,=\,{2\over g_{xx}}\,\Big[\,g^{tt}\,(\partial_t\,\phi)^2\,+\,
g^{xx}\,(\partial_3\,\phi)^2\,+\,g^{zz}\,(\partial_z\,\phi)^2\,\Big]\,\,,
\eeq
or, taking into account that $\partial_1\,\phi=\partial_2\,\phi=0$:
\beq
F^2\,=\,{2\over g_{xx}}\,g^{MN}\,\partial_{M}\,\phi\,\partial_{N}\,\phi\,\,.
\eeq
Thus, the action for the perturbation $\phi$ is:
\beq
S_{\phi}\,=\,-{1\over 16\pi G_N}\,
\int d^{d+1} x\, \sqrt{-g}\,\,{1\over 2}\,
g^{MN}\,\partial_{M}\,\phi\,\partial_{N}\,\phi\,\,,
\eeq
which is the canonical form of the action for a scalar field, with normalization constant
\beq
q\,=\,16\pi \,G_N\,\,.
\eeq
It follows from our general calculation of section \ref{transport_coefficients} of the transport coefficients  for a scalar field that the shear viscosity is given by:
\beq
\eta\,=\,{1\over 16\pi \,G_N} {A_H\over V}\,\,,
\eeq
and, therefore, the ratio $\eta/s$ is just \cite{Kovtun:2004de,Son:2007vk}:
\beq
{\eta\over s}\,=\,{1\over 4\pi}\,\,.
\eeq
In ordinary units the ratio $\eta/s$ is given by:
\beq
{\eta\over s}\,=\,{\hbar\over 4\pi k_B}\,\,,
\eeq
where $k_B$ is the Boltzmann constant. Notice that this result does not depend on the metric chosen. It is valid for any theory with a gravity dual given by Einstein gravity coupled to matter fields. In this sense it is a universal result valid at $\lambda\to\infty$ (infinite coupling limit). The finite coupling corrections can also be calculated. For ${\cal N}=4$ SYM one gets:
\beq
{\eta\over s}\,=\,{1\over 4\pi}\Big(1+{15\,\zeta(3)\over \lambda^{{3\over 2}}}+\cdots\Big)\,\,,
\eeq
where $\zeta(x)$ is the Riemann zeta function ($\zeta(3)=1.2020$). In general, $\eta/s$ at $\lambda\to\infty$ is very small:
\beq
{\eta\over s}\,=\,0.07957\,\,.
\eeq
The finite coupling corrections make $\eta/s$ increase. It is interesting to compare with the  weak coupling calculation, valid when $\lambda\to 0$:
\beq
{\eta\over s}\,=\,{A\over \lambda^2\,\log\Big({B\over \sqrt{\lambda}}\Big)}\,\,,
\label{eta/s_perturbative}
\eeq
where $A$ and $B$ are constant coefficients that depend on the theory. 
Notice that  in (\ref{eta/s_perturbative}) $\eta/s\to\infty$ as $\lambda\to 0$. This is because a weakly coupled gauge theory is a gas with strong dissipative effects, in which momentum can be transported over long distances due to the long free path. In contrast a strongly coupled plasma is an almost perfect fluid in which momentum is rapidly transferred between layers of sheared fluid. 

Kovtun, Son and Starinets (KSS) \cite{Kovtun:2004de} conjectured that $1/ (4\pi)$ is the lower bound for $\eta/s$. The lowest values of the  $\eta/s$ ratio found experimentally occur in two physical systems: the quark-gluon plasma created in heavy ion collisions at RHIC and the  ultracold atomic Fermi gases at very low temperature.  Both systems have $\eta/s$ which is slightly above $1/(4\pi)$.

Changing the gravity theory the KSS bound is violated. For example, by adding higher curvature terms as in the Gauss-Bonnet gravity, whose action is given by:
\beq
S_{GB}\,=\,{1\over 16\pi G_5}\,\int d^5x \sqrt{-g}\,\Big[
R-2\Lambda\,-\,{3\over \Lambda}\,\lambda_{GB}\,
\Big(R^2\,-\,4R_{\mu\nu}R^{\mu\nu}\,+\,R_{\mu\nu\rho\sigma}
R^{\mu\nu\rho\sigma}\Big)\Big]\,\,.
\eeq
For this gravity theory,  the following $\eta/s$ ratio is obtained:
\beq
{\eta\over s}\,=\,{1-4\lambda_{GB}\over 4\pi}\,\,,
\eeq
which violates the KSS bound if $\lambda_{GB}>0$.

\section*{Final remarks}

In these lecture notes we have reviewed the basic features of the AdS/CFT duality, focusing on its conceptual foundations and on some particular applications. We have only scratched the surface of the subject, which  in the last years has become a highly diversified field with many ramifications and connections. For the reader interested in knowing more on some of these applications of the holographic duality, let us quote some review articles, where the reader can find detailed accounts and the original references.  

In this review we have mostly dealt with the gravitational description of 
${\cal N}=4$ $SU(N)$ gauge theory, which only contains fields transforming in the adjoint representation of the gauge group. In this sense ${\cal N}=4$ SYM is a theory of pure glue. In order to extend the duality to theories closer to particle physics phenomenology one should be able to include flavor fields transforming in the fundamental representation of the gauge group (\ie\ quarks). This can be done by adding the so-called flavor branes, as reviewed  in \cite{CasalderreySolana:2011us,Erdmenger:2007cm}.  Moreover, the gauge/gravity duality can be extended to include less supersymmetric theories exhibiting confinement (see \cite{Edelstein:2006kw}) and one can construct holographic duals of quantum cromodynamics (\cite{CasalderreySolana:2011us,Kim:2012ey,Peeters:2007ab,Mateos:2007ay}). The holographic methods can also be used to study dynamical electroweak symmetry breaking, in the framework of walking technicolor models \cite{Piai:2010ma}. On the other hand, the AdS/CFT correspondence has unveiled the integrable character of planar ${\cal N}=4$ SYM and has allowed to extend the duality beyond the supergravity regime \cite{Beisert:2010jr}. 

One of the more interesting  recent developments of the gauge/gravity duality is the application of the string theoretical ideas to the down-to-earth problems of condensed matter physics. Indeed, condensed matter physics is full of strongly-coupled systems which display quantum criticality with specific scaling laws. The gauge/gravity duality allows to map this scaling behavior to the general covariance of a gravity theory, for which one can apply the calculational tools and physical intuition of general relativity. 
In this way one can  model the behavior of unusual phases of matter, such us strange metals or unconventional superconductors \cite{Hartnoll:2009sz, Sachdev:2010ch,Sachdev:2011wg,Green:2013fqa}. In this context holography is emerging as a new tool to understand the collective quantum behavior not explained by the conventional paradigms, such as the Fermi-liquid theory. 

It is also interesting to point out the connection between the AdS/CFT correspondence and quantum information theory. In particular, there is a holographic proposal for the entanglement entropy \cite{Nishioka:2009un,Takayanagi:2012kg}, which allows a simple geometrical calculation of the latter as the area of a minimal surface. Let us finally mention that holography has also been applied to the study of strongly-coupled hydrodynamics. This particular version of the duality is called the fluid-gravity correspondence \cite{Rangamani:2009xk,Hubeny:2011hd}.

Hopefully these lectures will stimulate the reader to explore some of the topics listed above. 

\section*{Acknowledgments}

I am grateful to Yago Bea,  Niko Jokela,  Javier Mas and Ricardo V\'azquez for their comments and help in the preparation of these lecture notes. I also thank Carlos Merino for his invitation to deliver the course on the AdS/CFT correspondence at the third IDPASC school.  This  work is funded in part by the Spanish grant 
FPA2011-22594,  by Xunta de Galicia (Conseller{\'i}a de Educaci\'on, grant 
INCITE09 206 121 PR and grant PGIDIT10PXIB206075PR),  by the 
Consolider-Ingenio 2010 Programme CPAN (CSD2007-00042), and by FEDER.

%
%
%

\end{document}